\documentclass[aps,prf,superscriptaddress,longbibliography]{revtex4-1}

\usepackage{amsmath,rotating}
\usepackage{dashrule}

\usepackage{psfrag}
\usepackage{latexsym}
\usepackage{longtable}
\usepackage{mathrsfs}
\usepackage{graphicx}
\usepackage{pifont}
\usepackage{amsmath}
\usepackage{subfigure}
\usepackage{mathtools}
\usepackage{mathcomp}
\usepackage{mathdots}
\usepackage{rotating}
\usepackage{array}
\usepackage{color}
\usepackage{algorithm}
\usepackage{algorithmic}
\usepackage[normalem]{ulem}

\usepackage{amssymb}
\usepackage[dvipsnames]{xcolor}

\usepackage{tikz}
\usetikzlibrary{arrows,shapes,trees,calc,positioning}
\usetikzlibrary{matrix}

\newcommand{\bq}{\bf{q}}

\newtheorem{remark}{Remark}

\newcommand{\beq}{\begin{equation}}
\newcommand{\eeq}{\end{equation}}
\newcommand{\bseq}{\begin{subequations}}
\newcommand{\eseq}{\end{subequations}}
\newcommand{\beqn}{\begin{eqnarray}}
\newcommand{\eeqn}{\end{eqnarray}}
\newcommand{\ba}{\begin{array}}
\newcommand{\ea}{\end{array}}
\newcommand{\bct}{\begin{center}}
\newcommand{\ect}{\end{center}}
\newcommand{\btmz}{\begin{itemize}}
\newcommand{\etmz}{\end{itemize}}
\newcommand{\benum}{\begin{enumerate}}
\newcommand{\eenum}{\end{enumerate}}

\newcommand{\bv}{{\bf v}}

\newcommand{\be}{\begin{equation}}
\newcommand{\ee}{\end{equation}}

\newcommand{\cplxs}{ C\kern -.35em \rule{0.03 em}{.7 ex}~   }

\def\complex{\hbox{C\kern -.45em \rule{0.03 em}{1.5 ex}}~}

\newcommand{\bi}{\begin{itemize}}
\newcommand{\ei}{\end{itemize}}

\newcommand{\bu}{{\bf u}}

\newcommand{\btab}{\begin{tabular}}
\newcommand{\etab}{\end{tabular}}

\newcommand{\non}{\nonumber}
\newcommand{\mrd}{\mathrm{d}}
\newcommand{\mre}{\mathrm{e}}
\newcommand{\mri}{\mathrm{i}}

\newcommand{\ds}{\displaystyle}

\newcommand{\DefinedAs}[0]{\mathrel{\mathop:}=}

\definecolor{bgblue}{rgb}{0.04,0.19,0.53}
\definecolor{dblue1}{rgb}{0,0.3,0.7}
\definecolor{dred}{rgb}{0.4,0.2,0}

\tikzstyle{block} = [draw,rectangle,thick,minimum height=2em,minimum width=1.0cm,
                                top color=blue!10, bottom color=blue!10]%
\tikzstyle{sum} = [draw,circle,inner sep=0mm,minimum size=2mm]%
\tikzstyle{connector} = [->,thick]%
\tikzstyle{line} = [thick]%
\tikzstyle{branch} = [circle,inner sep=0pt,minimum size=1mm,fill=black,draw=black]%
\tikzstyle{guide} = []%
\tikzstyle{deltablock} = [block, top color=red!10, bottom color=red!10]%
\tikzstyle{controlblock} = [block, top color=green!10, bottom color=green!10]%
\tikzstyle{weightblock} = [block, top color=orange!10, bottom color=orange!10]%
\tikzstyle{clpblock} = [block, top color=cyan!10, bottom color=cyan!10]%
\tikzstyle{block_dim} = [draw,rectangle,thick,minimum height=2em,minimum width=2em,
                                color=black!15]%
\tikzstyle{sum_dim} = [draw,circle,inner sep=0mm,minimum size=2mm,color=black!15]%
\tikzstyle{connector_dim} = [->,thick,color=black!15]%
\tikzstyle{smalllabel} = [font=\footnotesize]%
\tikzstyle{axiswidth}=[semithick]%
\tikzstyle{axiscolor}=[color=black!50]%
\tikzstyle{help lines} =[color=blue!40,very thin]%
\tikzstyle{axes} = [axiswidth,axiscolor,<->,smalllabel]%
\tikzstyle{axis} = [axiswidth,axiscolor,->,smalllabel]%
\tikzstyle{tickmark} = [thin,smalllabel]%
\tikzstyle{plain_axes} = [axiswidth,smalllabel]%
\tikzstyle{w_axes} = [axiswidth,->,smalllabel]%
\tikzstyle{m_axes} = [axiswidth,smalllabel]%
\tikzstyle{dataplot} = [thick]%
\begin{document}

\title{\Large \bf
Modeling mode interactions in boundary layer flows \\ via the Parabolized Floquet Equations}

\author{Wei Ran}
\email[E-mail:]{wran@usc.edu}
\affiliation{Department of Aerospace and Mechanical Engineering, University of Southern California, 90089, California, USA}
\author{Armin Zare}
\email[E-mail:]{armin.zare@usc.edu}
\affiliation{Ming Hsieh Department of Electrical and Computer Engineering, University of Southern California, 90089, California, USA}
\author{M.\ J.\ Philipp Hack}
\email[E-mail:]{mjph@stanford.edu}
\affiliation{Center for Turbulence Research, Stanford University, 94305, California, USA}
\author{Mihailo R.\ Jovanovi\'c}
\email[E-mail:]{mihailo@usc.edu}
\affiliation{Ming Hsieh Department of Electrical and Computer Engineering, University of Southern California, 90089, California, USA}

	\begin{abstract}
        In this paper, we develop a model based on successive linearization to study interactions between different modes in boundary layer flows. Our method consists of two steps. First, we augment the Blasius boundary layer profile with a disturbance field resulting from the linear Parabolized Stability Equations (PSE) to obtain the modified base flow; and, second,  we draw on Floquet decomposition to capture the effect of mode interactions on the spatial evolution of flow fluctuations via a sequence of linear progressions. The resulting Parabolized Floquet Equations (PFE) can be conveniently advanced downstream to examine the interaction between different modes in slowly varying shear flows. We apply our framework to two canonical settings of transition in boundary layers; the H-type transition scenario that is initiated by exponential instabilities, and streamwise elongated laminar streaks that are triggered by the lift-up mechanism.  We demonstrate that the PFE capture the growth of various harmonics and provide excellent agreement with the results obtained in direct numerical simulations and in experiments.
           \end{abstract}

\maketitle

\section{Introduction}
\label{sec.intro}

A thorough understanding of the mechanisms driving laminar-turbulent transition in boundary layer flows is crucial for the prediction of the point of transition and for the design of air and water vehicles. In the past thirty years, remarkable progress has been made on simulating the physics of transitional flows using models with various levels of fidelity. In spite of this, the multi-layer nature of transition and the inherent complexity of the Navier-Stokes (NS) equations have hindered the development of practical control strategies for delaying transition in boundary layer flows~\cite{kimbew07}. Direct Numerical Simulations (DNS) have opened the way to accurate investigations of the underlying physics of transitional flows~\cite{raimoi93,risfas95,wumoi09}. However, due to their high complexity and large number of degrees of freedom, nonlinear dynamical models that are based on the NS equations are not suitable for analysis, optimization, and control. On the other hand, nontrivial challenges, including lack of robustness, may arise in the model-based control of reduced-order models that are obtained using data-driven techniques~\cite{noamortad11}.

Linearization of the NS equations around the mean-velocity profile results in models that are well-suited for analysis and control synthesis using tools from modern robust control~\cite{kimbew07}. In particular, stochastically forced linearized NS equations have been used to capture structural and statistical features of transitional~\cite{farioa93,bamdah01,mj-phd04,jovbamJFM05} and turbulent~\cite{jovbamCDC01,hwacosJFM10b,moajovJFM12,zarjovgeoJFM17} channel flows. In these models, stochastic forcing may be utilized to model the impact of exogenous excitation sources or to capture the effect of nonlinear terms in the NS equations. Moreover, in conjunction with the parallel-flow assumption, the linearized NS equations are convenient for modal and non-modal stability analysis of spatially evolving flows~\cite{schhen01,sch07}. However, this approach does not account for the effect of the spatially evolving base flow on the stability of the boundary layer. Global stability analysis addresses this issue by accounting for the spatially varying nature of the base flow and discretizing all inhomogeneous spatial directions~\cite{ehrgal05,aakehrgalhen08,pargosthekim16}. Although accurate, this approach leads to problem sizes that may be prohibitively large for flow control and optimization.

In the boundary layer flow, primary disturbances are instigated via receptivity processes that involve internal or external perturbations~\cite{kac94}, e.g., acoustic noise, free-stream turbulence, and surface roughness. Depending on the amplitude of these disturbances, the transition process may take various paths to breakdown~\cite{morresher94,sarreeker02}. In particular, primary disturbances can be amplified through modal instability mechanisms or they may experience non-modal amplification, e.g., via transient growth, the lift-up~\cite{lan75,lan80} and Orr mechanisms~\cite{orr1907,butfar92}. Both pathways can intensify disturbances beyond the critical threshold, trigger secondary instabilities, and induce a strong energy transfer from the mean flow into secondary modes~\cite{her88}. The H-type~\cite{kaclev84,her88,sayhammoi13} and K-type~\cite{kletidsar62,sayhammoi13} transition scenarios are typical cases that are triggered by secondary instability mechanisms. Such mechanisms have also been shown to play an important role in the breakdown of laminar streaks at the later stages of transition~\cite{andbrabothen01,asaminnis02,brahen02,frabratalcos04,haczak16}. All of these are initiated after the significant growth of the primary disturbances which intensify the role of nonlinear interactions. The modulation of the base flow by the primary perturbations precludes the usual normal-mode assumptions made in the derivation of the Orr-Sommerfeld equation. Instead, the physics of these secondary growth mechanisms have been commonly studied using Floquet analysis~\cite{her88,andbrabothen01,bra-phd03} and the parabolized stability equations (PSE)~\cite{josstrcha93,her94,her97}.

The PSE were introduced to account for non-parallel and nonlinear effects and thereby overcome challenges associated with analyses based on a parallel-flow assumption. In particular, the PSE were developed as a means to refine predictions of parallel flow analysis in slowly varying flows~\cite{berherspa92,her94,her97}, e.g., in the laminar boundary layer. The PSE have also been adapted to account for the dynamics of three-dimensional flows that depend strongly on two spatial directions~\cite{galhal05,detparsanthe13,parhanthehen15}, and more recently, they have been used to model the amplification of disturbances in DNS and wall-modeled large-eddy simulation of transitional boundary layers~\cite{lozhacmoi18}. In spite of these successes, the nonlinear nature of this framework has hindered their utility in systematic optimal flow control design.
In general, the linear PSE provide reasonable predictions for the evolution of individual primary modes such as Tollmien-Schlichting (TS) waves~\cite{her94}. Moreover, the predictive capability of the linear PSE has been further refined by modeling the effect of nonlinear terms as a stochastic source of excitation~\cite{ranzarhacjovACC17}. However, secondary growth mechanisms that lead to laminar-turbulent transition of the boundary layer flow originate from interactions between different modes and these interactions cannot be explicitly accounted for using such techniques.

In the transitional boundary layer, primary instability mechanisms can cause disturbances to grow to finite amplitudes and saturate at steady or quasi-steady states. Floquet stability analysis identifies secondary instability modes as the eigen-modes of the NS equations linearized around the modified base flow profile that contains spatially periodic primary velocity fluctuations. In the corresponding eigenvalue problem, the operators inherit a periodic structure from the underlying periodicity of the base flow and, as a result, capture primary-secondary mode interactions. Such representations that account for mode interactions also appear in the analysis of distributed systems with spatially or temporally periodic coefficients~\cite{farjovbam08,jovfarAUT08} as well as in the model-based design of sensor-free periodic strategies for controlling transitional and turbulent wall-bounded shear flows~\cite{jovPOF08,moajovJFM10,liemoajovJFM10,moajovJFM12}.

In this paper, we take a step toward developing low-complexity models that account for harmonic interactions via a linear progression. To capture the dominant mode interactions while taking into account non-parallel effects, we introduce a computational framework that combines concepts from Floquet analysis and the linear PSE. The resulting equations are advanced downstream via a marching procedure. Our framework thus inherits the ability to account for mode interactions from Floquet theory while maintaining the low-complexity of the linear PSE. As a result, the proposed approach not only captures the essential physics of transitional boundary layer flows, but also opens the door to model-based control design.

Our presentation is organized as follows. In Sec.~\ref{sec.formulation}, we describe the linearized NS equations and the linear PSE. In Sec.~\ref{sec.PFE}, we derive the proposed Parabolized Floquet Equations (PFE) and explain the key features of our modeling framework. In Sec.~\ref{sec.H-type-transition}, we examine the growth of subharmonic modes in a typical H-type transition scenario and, in Sec.~\ref{sec.Nlstreak}, we employ our framework to study the formation of streaks in the boundary layer flow. We conclude the paper with remarks and outline of future research directions in Sec.~\ref{sec.conclusion}.

	\vspace*{-2ex}
\section{Background}
\label{sec.formulation}

We first present the equations that govern the dynamics of flow fluctuations in incompressible flows of Newtonian fluids and then provide details on our proposed model for the downstream marching of spatially growing fluctuations in the boundary layer flow.

\begin{figure}
\vspace{.3cm}
\begin{center}
         \includegraphics[width=8.5cm]
	{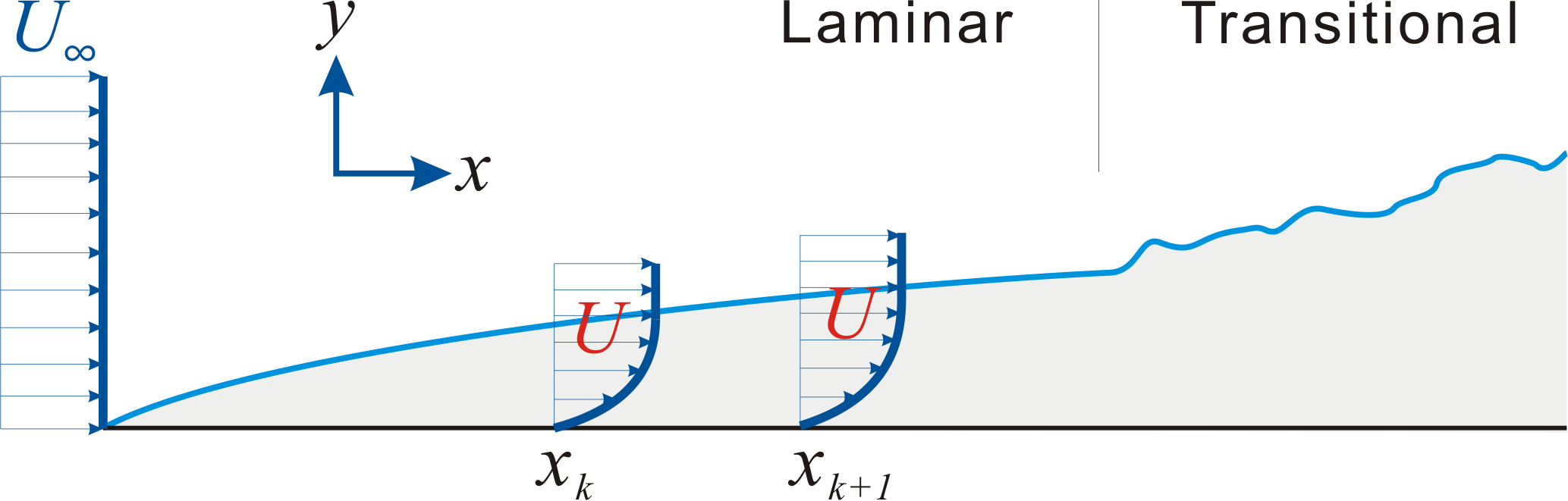}
\end{center}
\caption{Geometry of a transitional boundary layer flow.}
\label{fig.forcedBL}
\end{figure}

In a flat-plate boundary layer, with geometry shown in Fig.~\ref{fig.forcedBL}, the dynamics of infinitesimal fluctuations around a two-dimensional base flow $\bar{\bu} = [\,U(x,y)\,~V(x,y)\,~0\,]^T$ are governed by the linearized NS equations
\begin{align}
	\label{eq.turblin}
	\ba{rcl}
    		\bv_t
    		&\; = \;&
    		-
    		\left( \nabla \cdot \bar{\bu} \right) \bv
     		\; - \;
    		\left( \nabla \cdot \bv \right) \bar{\bu}
    		\; - \;
    		\nabla p
    		\; + \,
    		\dfrac{1}{Re_0}\,  \Delta \bv
    		\\[.15cm]
    		0
    		&\; = \;&
    		\nabla \cdot \bv,
	\ea
\end{align}
where $\bv = [\,u\,~v\,~w\,]^T$ is the vector of velocity fluctuations, $p$ denotes pressure fluctuations, $u$, $v$, and $w$ are the streamwise ($x$), wall-normal ($y$), and spanwise ($z$) components of the fluctuating velocity field, and $Re_0$ is the Reynolds number at the inflow location $x_0$. The Reynolds number is defined as $Re = U_\infty \delta/\nu$, where $\delta=\sqrt{\nu x/U_{\infty}}$ is the Blasius length scale at the streamwise location $x$, $U_\infty$ is the free-stream velocity, and $\nu$ is the kinematic viscosity. Spatial coordinates are non-dimensionalized by the Blasius length scale $\delta_0$ at the inflow location $x_0$, velocities by $U_\infty$, time by $\delta_0/U_\infty$, and pressure by $\rho U_\infty^2$, where $\rho$ is the fluid density.

It is customary to use the parallel-flow approximation to study the local stability of boundary layer flows to small amplitude disturbances~\cite{schhen01}. This approximation, in conjunction with Floquet theory, has also been used to investigate secondary instabilities that inflict transition~\cite{her88,schhen01}. However, the parallel-flow approximation excludes the effect of the evolution of the base flow on the amplification of disturbances. This issue can be addressed via global stability analysis which accounts for the spatially varying nature of the base flow by discretizing all inhomogeneous directions. Nevertheless, global analysis of spatially-evolving flows may be prohibitively expensive for analysis, optimization, and control purposes.

The PSE provide a computationally attractive framework for the spatial evolution of perturbations in non-parallel and weakly nonlinear scenarios~\cite{berherspa92,her94,her97}. They are obtained by removing terms of $O(1/Re^2)$ and higher from the NS equations and are significantly more efficient than conventional flow simulations based on the governing equations. In weakly non-parallel flows, e.g., in the pre-transitional boundary layer, flow fluctuations can be separated into slowly and rapidly varying components via the following decomposition for the fluctuation field ${\bq} = [\,u\,~v\,~w\,~p\,]^T$ in~\eqref{eq.turblin}. For a specific spanwise wavenumber and temporal frequency pair $(\beta,\omega)$, we consider
\begin{align}
	\label{eq.PSEansatz}
	\ba{rcl}
	{\bq}(x,y,z,t)
	& \;=\; &
	{\hat{\bq}}(x,y)\, \chi(x,z,t) ~+~ \mathrm{complex~conjugate},
	\\[0.25cm]
	\chi(x,z,t)
	& \;=\; &
	\exp \left( \mri \left(\alpha(x)\,x \,+\, \beta\, z \,-\, \omega\, t \right) \right),
	\ea
\end{align}
where $\hat{\bq}(x,y)$ and $\chi(x,z,t)$ are the shape and phase functions, and $\alpha(x)$ is the streamwise varying generalization of the wavenumber~\cite{berherspa92}. This decomposition separates slowly ($\hat{\bq}(x,y)$) and rapidly ($\chi(x,z,t)$) varying scales in the streamwise direction. The ansatz in Eq.~\eqref{eq.PSEansatz} provides a representation of oscillatory instability waves such as TS waves.

The ambiguity arising from the streamwise variations of both $\hat{\bq}$ and $\alpha$ is resolved by imposing the condition
	\[
	\int_{\Omega_y} \hat{\bq}^* \hat{\bq}_x \, \mrd y = 0,
	\]
where $\hat{\bq}^*$ denotes the complex conjugate transpose of the vector $\hat{\bq}$. In practice, this condition is enforced through the iterative adjustment of the streamwise wavenumber~\cite[Sec.~3.2.5]{her94}. Following the slow-fast decomposition highlighted in~\eqref{eq.PSEansatz}, the linearized NS equations are parabolized under the assumption that the streamwise variation of $\hat{\bq}$ and $\alpha$ are sufficiently small to neglect $\hat{\bq}_{xx}$, $\alpha_{xx}$, $\alpha_x \hat{\bq}_x$, $\alpha_x/Re_0$, and their higher order derivates with respect to $x$, resulting in the removal of the dominant ellipticity in the NS equations. The linear PSE take the form
\be
	\label{eq.lin-PSE}
    \mathbf{L}\,\hat{\bq}
	\;+\;
	\mathbf{M}\,\hat{\bq}_x
	~=~
	0,
\ee
where expressions for the operator-valued matrices ${\bf L}$ and ${\bf M}$ can be found in~\cite{her94}.

We next propose a two-step modeling procedure to study the dominant mode interactions in weakly-nonlinear mechanisms that arise in spatially evolving flows.

	\vspace*{-2ex}
\section{Parabolized Floquet equations}
\label{sec.PFE}

In the transitional boundary layer flow, primary instabilities can cause disturbances to grow to finite amplitudes and get saturated by nonlinearity. Secondary stability analysis examines the asymptotic growth of the resulting modulated state and is based on the linearized NS equations around the modified base flow
\begin{align}
	\label{eq.base-modulation}
        \bar{\bu}
        ~ =~
        \bu_0
        \;+\;
        \mathbf{u}_\mathrm{pr}.
\end{align}
Here, $\bu_0$ denotes the original base flow and $\mathbf{u}_\mathrm{pr}$ represents the primary disturbance field. Since $\bar{\bu}$ is typically spatially or temporally periodic, Floquet analysis is invoked to identify the spatial structure of exponentially growing fluctuations around $\bar{\bu}$. However, such analysis relies on a parallel flow assumption and it does not explicitly account for the spatially growing nature of the base flow. To account for the interactions of fluctuations with spatially growing modified base flow $\bar{\bu}$  in a computationally efficient manner, we introduce a framework which draws on Floquet theory to enhance the linear PSE. Our approach allows us to capture the dominant mode interactions in the fluctuating velocity field while accounting for non-parallel effects in the base flow.

Starting from a spatially or temporally periodic initial condition the linear PSE can be marched downstream to obtain the primary disturbance field. For example, such an initial condition can be obtained using stability analysis of the two-dimensional Orr-Sommerfeld equation or transient growth analysis of streamwise constant linearized equations (under the locally-parallel base flow assumption). When the periodic solutions to the linear PSE computation are superposed to the Blasius boundary layer profile, the modulated base flow~\eqref{eq.base-modulation} takes the following form
\be
	\label{eq.base-modulation-1}
	\bar{\bu}(x, y, z, t)
	~=~
	\ds{\sum_{m\,=\,-\infty}^{\infty} \bu_m(x, y)\, \phi_m(x,z,t)}.
\ee
Here, $\bu_0 (x,y) = [\,U_B(x,y)\,~V_B(x,y)\,~0\,]^T$ represents the Blasius boundary layer profile, $\phi_0 = 1$, $\bu_m$ and $\phi_m$ for $m \neq 0$ are the shape and phase functions corresponding to various harmonics that constitute flow structures of the primary disturbance field (such as TS waves or streaks), and $\bu_m^* = \bu_{-m}$. Note that each harmonic $\bu_m$ of the modified base flow $\bar{\bu}$ inherits a similar slow-fast structure from PSE (cf.~Eq.~\eqref{eq.PSEansatz}) in which the phase function $\phi_m$ is spanwise or streamwise/temporally periodic. For example, when TS waves are superposed to the Blasius boundary layer profile the phase functions $\phi_m$ are streamwise and temporally periodic; see Sec.~\ref{sec.H-type-transition} for details. The evolution of fluctuations around the modulated base flow~\eqref{eq.base-modulation-1} can be studied using the following expansion
\begin{align}
    \label{eq.general-ansatz}
	{\bq}(x,y,z,t)
	~=\;
	\sum_{n\,=\,-\infty}^{\infty} \hat{\bq}_n(x, y)\, \chi_n(x, z, t),
\end{align}
which, similar to PSE, involves a decomposition of disturbances into slowly ($\hat{\bq}_n$) and rapidly ($\chi_n$) varying components. Note that we follow classical Floquet decomposition~\cite{her84,her88} in assuming that the phase functions $\chi_n$ represent various harmonics of the same fundamental frequency/wavenumber as $\phi_m$ in Eq.~\eqref{eq.base-modulation-1}. As a result of this assumption the evolution of each harmonic mode in $\bq$ can contribute to the evolution of its neighboring harmonics via the periodicity of the modulated base flow~\eqref{eq.base-modulation-1}. For spanwise periodic modulations to the base flow, a concrete example of the form of the fluctuation field~\eqref{eq.general-ansatz} is discussed in Remark~\ref{rem-1}.

	\begin{remark}
	\label{rem-1}
When spanwise-periodic streaks with a fundamental wavenumber $\beta$ are superposed to the Blasius boundary layer profile, the modified base flow takes the form
\begin{align*}
	\bar{\bu}(x, y, z)
	~=~
	\ds{\sum_{m\,=\,-\infty}^{\infty} \bu_m(x, y)\, \mre^{\mri m\beta z}},
	\non
\end{align*}
and the spatial evolution of fluctuations that account for fundamental harmonics (in $z$) around this modulated base flow profile can be studied using the Fourier expansion
	\begin{align*}
	{\bq}(x, y, z)
	~=~
	\ds{\sum_{n \, = \, -\infty}^{\infty}\hat{\bq}_n(x,y)\, \mre^{\mri (\alpha_n (x) x \,+\, n \beta z)}}.
\end{align*}
Here, $\alpha_n(x)$ is the purely imaginary streamwise wavenumber of various harmonics, which can evolve in the streamwise direction similarly to linear PSE. Note that if $\alpha_n(x)$ is identical for all harmonics, we recover the nondispersive wavepacket assumed in Floquet stability analysis; see Sec.~\ref{sec.Nlstreak} for additional details.
	\end{remark}

Under the assumptions of linear PSE, the dynamics of fluctuations represented by~\eqref{eq.general-ansatz} can be studied using the Parabolized Floquet Equations (PFE)
\be
	\label{eq.PFE}
    	\mathbf{L}_F\, \hat{\bq}
	\;+\;
	\mathbf{M}_F\,\hat{\bq}_x
	~=~
	0.
\ee
The state in~\eqref{eq.PFE},
\begin{align*}
	\hat{\bq}
	~=~
	[ ~ \cdots~\;\hat{\bq}_{n-1}^T~\;\hat{\bq}_n^T~\;\hat{\bq}_{n+1}^T~\;\cdots ~ ]^T,
\end{align*}
contains all harmonics of $\bq$ in the periodic direction, i.e.,
\begin{align*}
	\hat{\bq}_n
	~=~
	[\;u_{n}^T~\;v_{n}^T~\;w_{n}^T~\;p_{n}^T\;]^T,
\end{align*}
and the operators $\mathbf{L}_F$ and $\mathbf{M}_F$ inherit the following bi-infinite structure from the periodicity of the phase functions $\phi_m$ in the modified base flow~\eqref{eq.base-modulation-1},
\begin{eqnarray}
	\label{eq.toep}
	\mathbf{L}_F
	~\DefinedAs~
	\left[
	\ba{ccccc}
         \ddots &\, \vdots &\, \vdots &\, \vdots &\, \rotatebox{90}{$\ddots$}
         \\[0.2cm]
         \cdots &\, \mathbf{L}_{n-1,0} &\, \mathbf{L}_{n-1,+1} &\, \mathbf{L}_{n-1,+2} &\, \cdots
         \\[0.2cm]
         \cdots &\, \mathbf{L}_{n,-1} &\, \mathbf{L}_{n,0} &\, \mathbf{L}_{n,+1} &\,  \cdots
         \\[0.2cm]
         \cdots &\,  \mathbf{L}_{n+1,-2} &\, \mathbf{L}_{n+1,-1} &\, \mathbf{L}_{n+1,0} &\, \cdots
         \\[0.2cm]
         \rotatebox{90}{$\ddots$} &\, \vdots &\, \vdots &\, \vdots &\, \ddots
    	\ea
	\right].
\end{eqnarray}
Note that the operator $\mathbf{L}_{i,j}$ captures the influence of the $j$th harmonic $\hat{\bq}_{j}$ on the dynamics of the $i$th harmonic $\hat{\bq}_{i}$. In practice, the generally bi-infinite structures of the state and operators in Eq.~\eqref{eq.PFE} are truncated to account for the spatial evolution of a finite number of {\em essential\/} modes.

In what follows, we employ the PFE to examine two canonical problems:
	\bi
 \item the H-type transition scenario (Sec.~\ref{sec.H-type-transition}); and
 \item the formation of streamwise elongated streaks in laminar boundary layer flow (Sec.~\ref{sec.Nlstreak}).
 \ei
In the wall-normal direction, homogenous Dirichlet boundary conditions are imposed,
\begin{align*}
	\ba{rclrclrcl}
	u_n(0) &=& 0,&\quad v_n(0) &=& 0, &\quad w_n(0) &=& 0
	\\[0.2cm]
	u_n(L_y) &=& 0,&\quad v_n(L_y) &=& 0, &\quad w_n(L_y) &=& 0
\ea
\end{align*}
where $L_y$ denotes the height of the computational domain. We discretize differential operators $\mathbf{L}_F$ and $\mathbf{M}_F$ using a pseudospectral scheme with $N_y$ Chebyshev collocation points in the wall-normal direction~\cite{weired00} and employ an implicit Euler method to march the PFE~\eqref{eq.PFE} in the streamwise direction with constant step-size $\Delta x$.

	\vspace*{-2ex}
\subsection{Two-step modeling procedure}
\label{sec.2stepprocedure}

\begin{figure}
	\begin{centering}
	\begin{tabular}{l}
	\subfigure[]{\label{fig.PFEdiagram}}
	 \\[-.4cm]
	\hspace{.6cm}
    \includegraphics[height=3.9cm]{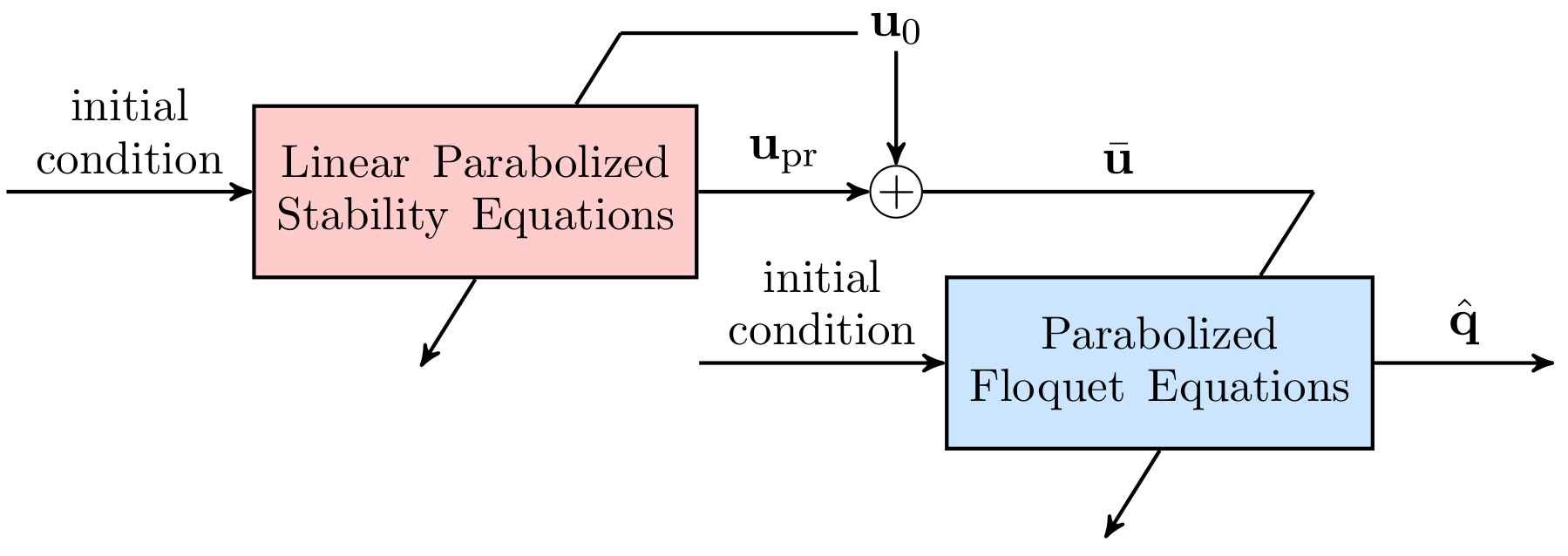}
		\\[-.2cm]
	\subfigure[]{\label{fig.equilibrium}}
	\\[-.4cm]
	\hspace{.6cm}
    \includegraphics[height=3cm]{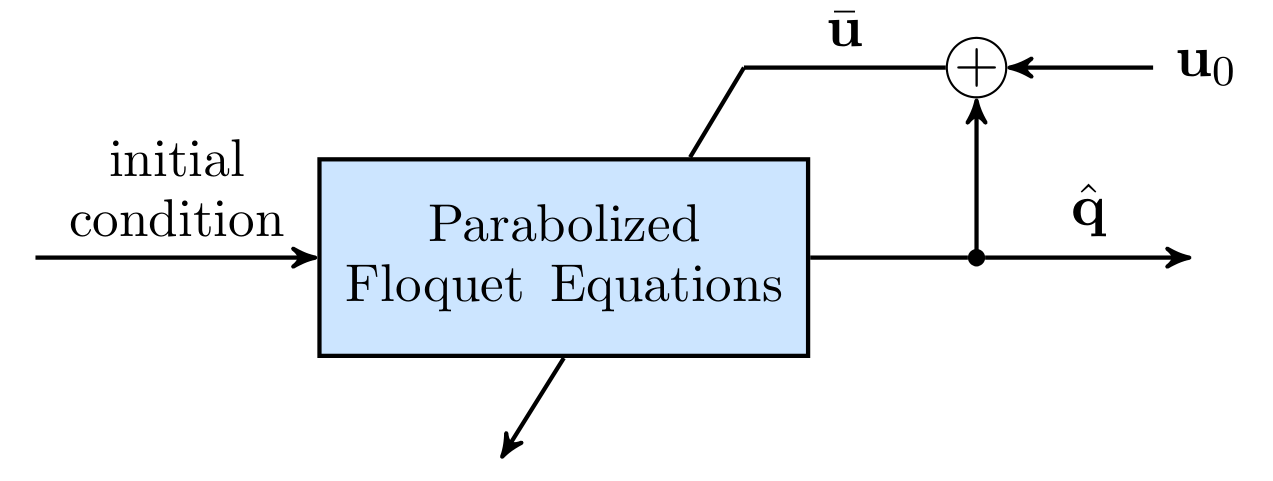}
		\end{tabular}
	\caption{(a) The PFE are triggered with a primary disturbance ${\bu}_{\mathrm{pr}}$ that results from linear PSE and modulates the base flow. The diagonal lines represent the base flows that enter as coefficients into the linear PSE and PFE, respectively. (b) When the secondary disturbances  {are of} the same frequencies and spatial wavenumbers as the primary disturbances, the dominant harmonics resulting from  {each PFE iteration (e.g., $\hat{\bq}_{0}$, $\hat{\bq}_{\pm1}$, and $\hat{\bq}_{\pm2}$ in the formation of streaks)} can be used to update the modulation to the base flow,  {iterate} the PFE, and compute an equilibrium configuration.}
	\label{fig.PFEdiagram-equilibrium}
	\end{centering}
\end{figure}

To model the effect of mode interactions in weakly nonlinear regimes we consider the following two-step procedure:
\begin{enumerate}
	\item{The linear PSE are used to march the primary harmonic and obtain the corresponding velocity profile ${\bu}_{\mathrm{pr}}$ at each streamwise location.}
	\item{The PFE are used to march all harmonics $\hat{\bq}$ and obtain the spatial evolution of velocity fluctuations around the modified base flow $\bar{\bu} = \bu_0
        +
        \mathbf{u}_\mathrm{pr}$.}
\end{enumerate}
The PFE are thus used to study the effect of dominant harmonic interactions on the growth of disturbances in the streamwise direction. The block diagram in Fig.~\ref{fig.PFEdiagram} illustrates our modeling procedure.

When the secondary disturbances contain the same temporal frequency and spatial wavenumbers as the primary disturbances, the dominant harmonics resulting from the PFE (e.g.,  $\hat{\bq}_{0}$, $\hat{\bq}_{\pm1}$, and $\hat{\bq}_{\pm2}$ in the formation of streaks) can be subsequently used to update the modulation to the base flow, iterate the PFE computation, and thereby provide an equilibrium configuration. The block diagram in Fig.~\ref{fig.equilibrium} illustrates how our framework can be employed to correct primary disturbances that subsequently modulate the base flow. This should be compared and contrasted to the conventional Floquet analysis in which the frequencies and wavenumbers of the identified secondary instabilities are different from those in the primary disturbances. Our computational experiments in Sec.~\ref{sec.Nlstreak} demonstrate that the flow state of the feedback interconnection in Fig.~\ref{fig.equilibrium} converges after a certain number of iterations. While the equilibrium configuration in Fig.~\ref{fig.equilibrium} implies that our framework is inherently nonlinear, each step in our iterative procedure is linear. We note that a similar iterative method was successfully utilized for model-based design of spanwise wall oscillations in turbulent channel flows~\cite{moajovJFM12}.

	\vspace*{-3ex}
\section{H-type transition}
\label{sec.H-type-transition}
	\vspace*{-1ex}

We next apply our approach to model an H-type transition scenario in a zero-pressure-gradient boundary layer~\cite{her88}. This route to transition begins with the exponential growth of two-dimensional TS waves which, upon reaching a critical amplitude, become unstable to secondary disturbances. The modulation of the Blasius profile by the TS waves induces the amplification of otherwise stable oblique modes which have half the frequency of the TS waves. While the linear PSE can be used to characterize the spatial evolution of modes arising from primary instabilities, secondary instabilities that trigger the growth of subharmonic modes call for an expansion in the harmonics of the modulated base flow. Such a growth mechanism cannot be identified via the normal-mode ansatz employed in the linear PSE but it can be captured by marching the PFE, a model resulting from a combination of linear PSE with Floquet decomposition.

	\vspace*{-2ex}	
\subsection{Setup}
\label{sec.H-type-Setup}

At the initial position $x_0$ and for a real-valued temporal frequency $\omega$, the fundamental mode in the H-type scenario is identified as the eigenvector corresponding to the most unstable complex eigenvalue $\alpha$ from the discrete spectrum of the standard two-dimensional Orr-Sommerfeld equation; see~\cite[Sec.~7.1.2]{schhen01} for additional details. The linear PSE can be used to march this fundamental mode, which is in the form of a two-dimensional TS wave, and obtain a reasonable prediction of its spatial growth~\cite{berherspa92}. We use the resulting solution to a primary linear PSE computation to augment the Blasius boundary layer profile $\bu_0$ in the base flow for the PFE as
\begin{align}
    \ba{rcl}
        U(x,y)
        &\;=\;&
        U_{B}(x,y)
        \;+\; U_{T}(x,y)\,\mre^{\mri \alpha_r (x\,-\,c\,t)} \;+\; U_{T}^*(x,y)\,\mre^{-\mri \alpha_r (x\,-\,c\,t)}
        \\[0.25cm]
        V(x,y)
        &\;=\;&
        V_{B}(x,y)
        \;+\; V_{T}(x,y)\,\mre^{\mri \alpha_r (x\,-\,c\,t)} \;+\; V_{T}^*(x,y)\,\mre^{-\mri \alpha_r (x\,-\,c\,t)}
        \\[0.25cm]
        W(x,y)
        &\;=\;&
        0.
        \ea
        \label{eq.baseflow-Htype}
\end{align}
Equivalently, the base flow $\bar{\bu}$ can be written in the following compact form
\[
	\label{eq.base-floquet-Htype}
	\bar{\bu}(x, y, t)
	~=~
	\ds{\sum_{m\,=\,-1}^{1} \bu_m(x, y)\, \mre^{\mri m \alpha_r (x\,-\,c\,t)}},
\]
where $\bu_0$ is the Blasius profile. In Eq.~\eqref{eq.baseflow-Htype}, $\alpha_r$ is the real part of the prescribed streamwise wavenumber of the fundamental mode  $\alpha(x)$ resulting from linear PSE, $c=\omega/\alpha_r$ denotes the phase speed of the fundamental and subharmonic modes in the fixed (laboratory) frame, and $[\,U_T(x,y)\,~V_T(x,y)\,~0\,]^T$ denotes the TS wave whose local amplitude and shape are obtained from the linear PSE computation.  Note that the exponential growth resulting from the imaginary part of the wavenumber, $\alpha_i$, is absorbed into the amplitudes of $U_T(x,y)$ and $V_T(x,y)$.

To study the evolution of subharmonic modes that are triggered via secondary instability mechanisms, we follow the Floquet decomposition which was originally conducted in the moving frame~\cite[Sec.~8.2]{schhen01} using the following Fourier expansion, but in the fixed (laboratory) frame
\be
	{\bq}(x,y,z,t)
	~=~
	\mre^{\mri \beta z}
	\ds{\sum_{n\,=\,-\infty}^{\infty}}
	\hat{\bq}_n(x,y)\, \mre^{\mri\, (n\,+\,0.5)\,\alpha_r (x\,-\,c\,t) \,+\, \mri\, \gamma_n x \,-\, \mri\,\sigma_n t}
    \label{eq.H-type-ansatz-orig}
\ee
where $\hat{\bq}_n^* = \hat{\bq}_{-n-1}$, and $\gamma_n$ and $\sigma_n$ are the spatial and temporal detuning factors corresponding to the $n$th subharmonic. The imaginary and real parts of $\gamma_n$ ($\sigma_n$) denote the spatial (temporal) growth rate and the detuning in the wavenumber (frequency), respectively. In practice, the detuning factors of the wavenumber and frequency are negligible, i.e., $\gamma_n$ and $\sigma_n$ can be assumed to be purely imaginary. A further assumption of nondispersive wavepackets in accordance with classical Floquet analysis~\cite{her88} brings the ansatz for the fluctuation field to the following form
\be
    \ba{rcl}
	{\bq}(x,y,z,t)
	& = &
	\mre^{\mri\, \left(\beta z \,+\, c\,\gamma t \right)}
	\ds{\sum_{n\,=\,-\infty}^{\infty}}
	\hat{\bq}_n(x,y)\, \mre^{\mri\, \left[(n\,+\,0.5)\,\alpha_r \,+\, \gamma\right](x\,-\,c\,t)}
	\\[0.5cm]
	& = &
	\mre^{\mri\, \left(\beta z \,+\, \gamma x \right)}
	\ds{\sum_{n\,=\,-\infty}^{\infty}}
	\hat{\bq}_n(x,y)\, \mre^{\mri (n\,+\,0.5) \alpha_r (x\,-\,c\,t)},
    \ea
    \label{eq.H-type-ansatz}
\ee
where Gaster's transformation~\cite{gas62} has been used to replace $\sigma$ with $-c \gamma$ and all subharmonic modes are assumed to share a uniform detuning parameter $\gamma$. In making these approximations, we have followed~\cite{her88} in assuming that all Fourier components have the same phase speed, which is consistent with experimental studies~\cite{kaclev84}. While $\alpha_r$ in the PFE computation is prescribed by the solution of linear PSE for the primary disturbance field, we update the spatial growth rate $-\gamma$ via a similar scheme to the one used for the streamwise wavenumber update in PSE~\cite[Sec.~3.2.5]{her94}.

The PFE account for the interaction between different subharmonics by leveraging the slow-fast decomposition inherited from the solution of the linear PSE. By substituting the ansatz~\eqref{eq.H-type-ansatz} and the modulated base flow~\eqref{eq.baseflow-Htype} into the linearized NS equations, we arrive at the PFE which take the form of Eq.~\eqref{eq.PFE}. For this case study, the operators ${\bf L}_F$ and ${\bf M}_F$ in the PFE~\eqref{eq.PFE} are provided in Appendix~\ref{sec.appendix2}.

The procedure for obtaining the results presented in the next subsection can be summarized as follows:
\begin{enumerate}
  \item Solve the spatial eigenvalue problem corresponding to the Orr-Sommerfeld equations to obtain the initial complex wavenumber $\alpha(x_0)$ and the shape of initial TS wave $\mathbf{u}_\mathrm{pr}(x_0)$.
  \item Use linear PSE to march the TS wave downstream and obtain $\alpha(x)$ and $\mathbf{u}_\mathrm{pr}(x)$.
  \item Augment the Blasius boundary layer profile with the TS wave $\mathbf{u}_\mathrm{pr}(x)$ to obtain the modulated base flow $\bar{\bu}$.
  \item Obtain the initial uniform growth rate $\gamma$ and shape function $\hat{\bq}$ as the most unstable eigenvalue and eigenvectors from standard Floquet analysis~\cite{her88} at $x_0$.
  \item Use PFE to compute the spatial evolution of all subharmonic modes around the modulated base flow.
\end{enumerate}

	\vspace*{-2ex}
\subsection{Growth of subharmonic secondary instabilities}
\label{sec.H-type-growth}

We next examine the interaction of TS waves with subharmonic secondary instabilities in the classical H-type transition scenario. This problem was initially studied in~\cite{her84} and has been further explored using both experiments~\cite{kaclev84} and numerical simulations~\cite{josstrcha93,sayhammoi13}.

Following Refs.~\cite{kaclev84, her88}, the primary disturbance field is generated by marching a TS wave with a root-mean-square (rms) amplitude of $4.8\times 10^{-3}$ and frequency $\omega=0.496$ from $Re_0=424$ to $Re=700$ using linear PSE. We subsequently initialize the PFE with the most unstable eigen-mode from the classical Floquet analysis~\cite{her88} to study the growth of subharmonic secondary instabilities triggered by the TS wave. The initial rms amplitude of the subharmonic mode is $1.46\times 10^{-5}$ and its frequency and spanwise wavenumber are $\omega=0.248$ and $\beta=0.132$, respectively. The initial spatial growth rate  $- \gamma$ is obtained by applying Gaster's transformation to the temporal growth rate. We consider a truncation of the bi-infinite state ${\bq}$ with $2N$ modes, i.e., $n=-N \cdots N-1$, and a computational domain with $L_x\times L_y= 1100 \times 40$. Our computations demonstrate that {$N_y=80$, $\Delta x=15$,} and $N=2$ (i.e., $4$ subharmonic modes), provide sufficient accuracy in capturing the physics of H-type transition; see Appendix~\ref{sec.convergence} for a discussion on wall-normal grid-convergence and about the influence that the number of subharmonics has on our results.

\begin{figure}[!ht]
                \begin{tabular}{lcl}
        \subfigure[]{\label{fig.om0p0248subharmonicamp}}
        &&
        \subfigure[]{\label{fig.om0p0248primary20mode}}
        \\[-.4cm]
        \begin{tabular}{c}
                $u_{\mathrm{rms}}$
                \\
                \includegraphics[width=7.5cm]{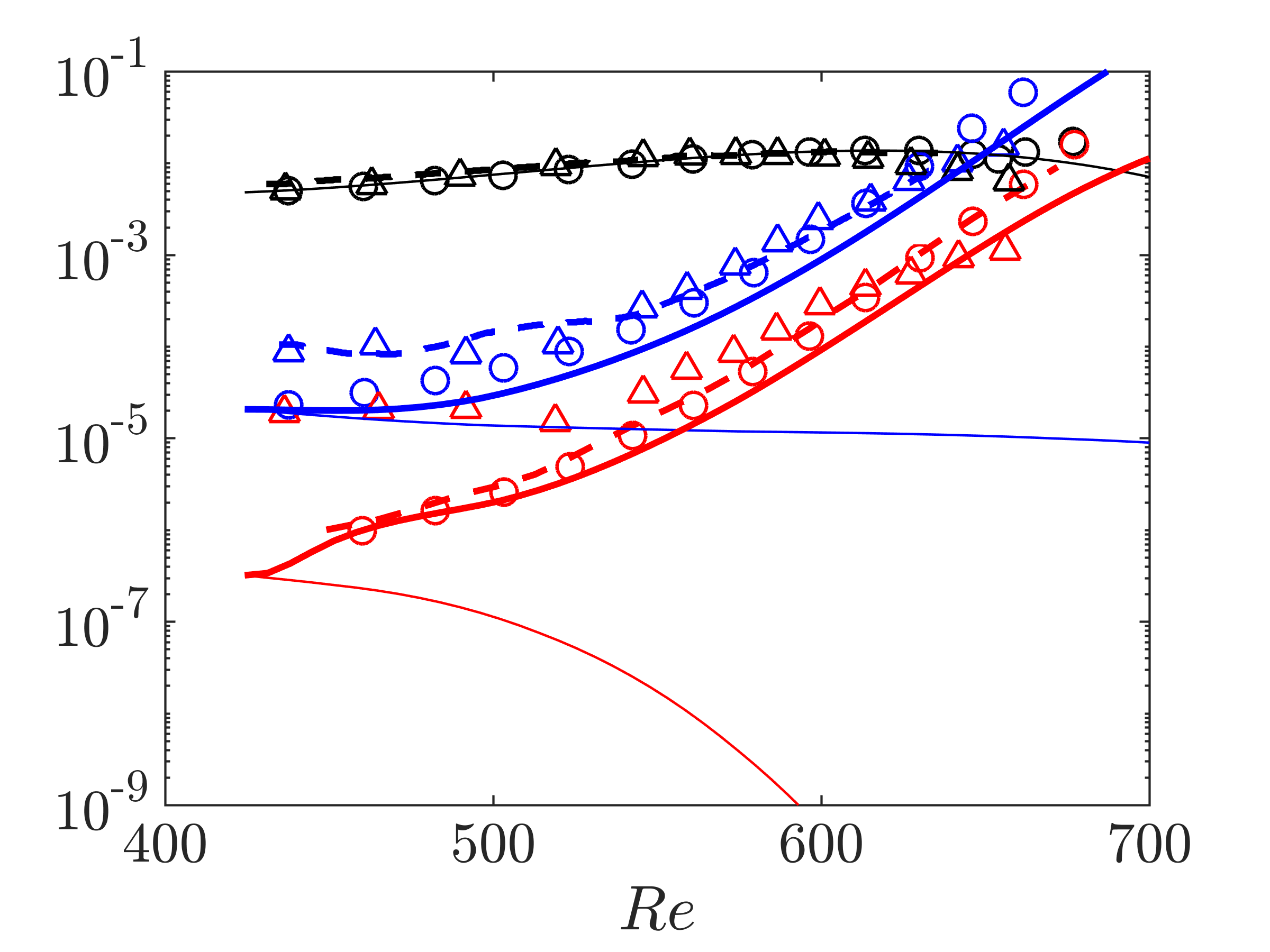}
        \end{tabular}
        &&
        \begin{tabular}{c}
                fundamental harmonic; ($2,0$) mode
                \\
                \includegraphics[width=7.5cm]{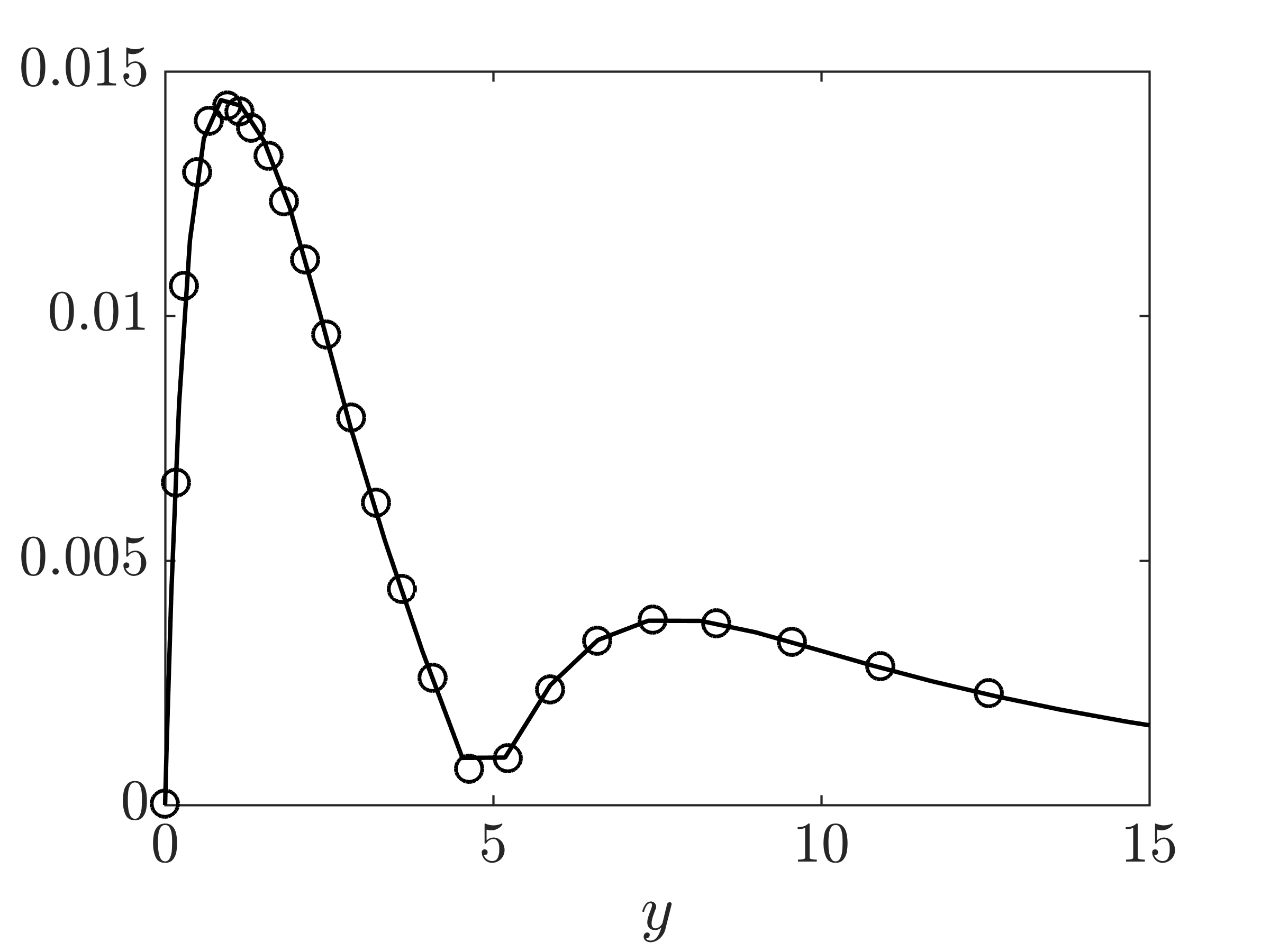}
        \end{tabular}
        \\[.1cm]
        \subfigure[]{\label{fig.om0p0248subharmonic11mode}}
        &&
        \subfigure[]{\label{fig.om0p0248subharmonic31mode}}
        \\[-.4cm]
        \begin{tabular}{c}
                ($1,1$) subharmonic
                \\
                \includegraphics[width=7.5cm]{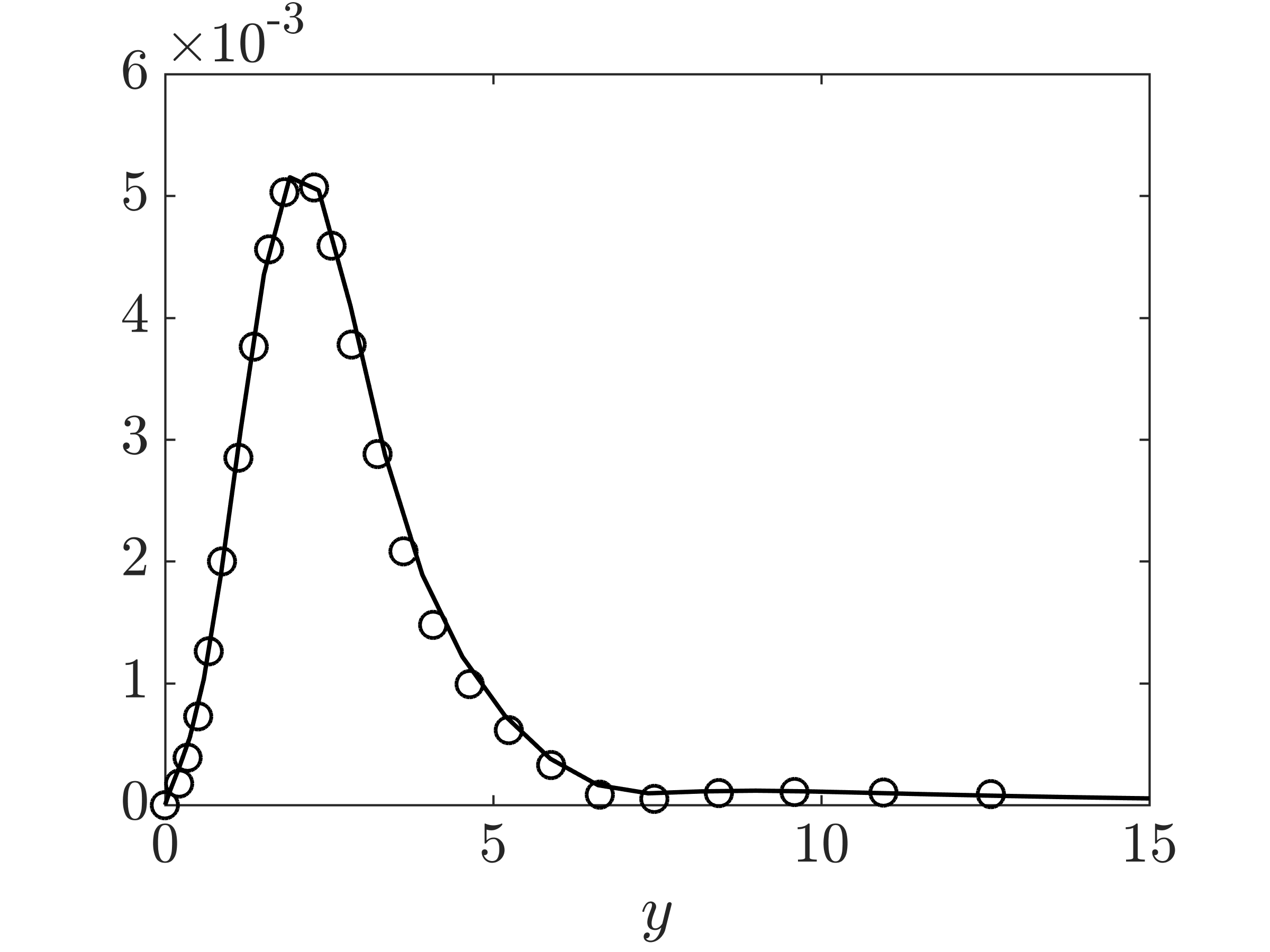}
        \end{tabular}
        &&
        \begin{tabular}{c}
                ($3,1$) subharmonic
                \\
                \includegraphics[width=7.5cm]{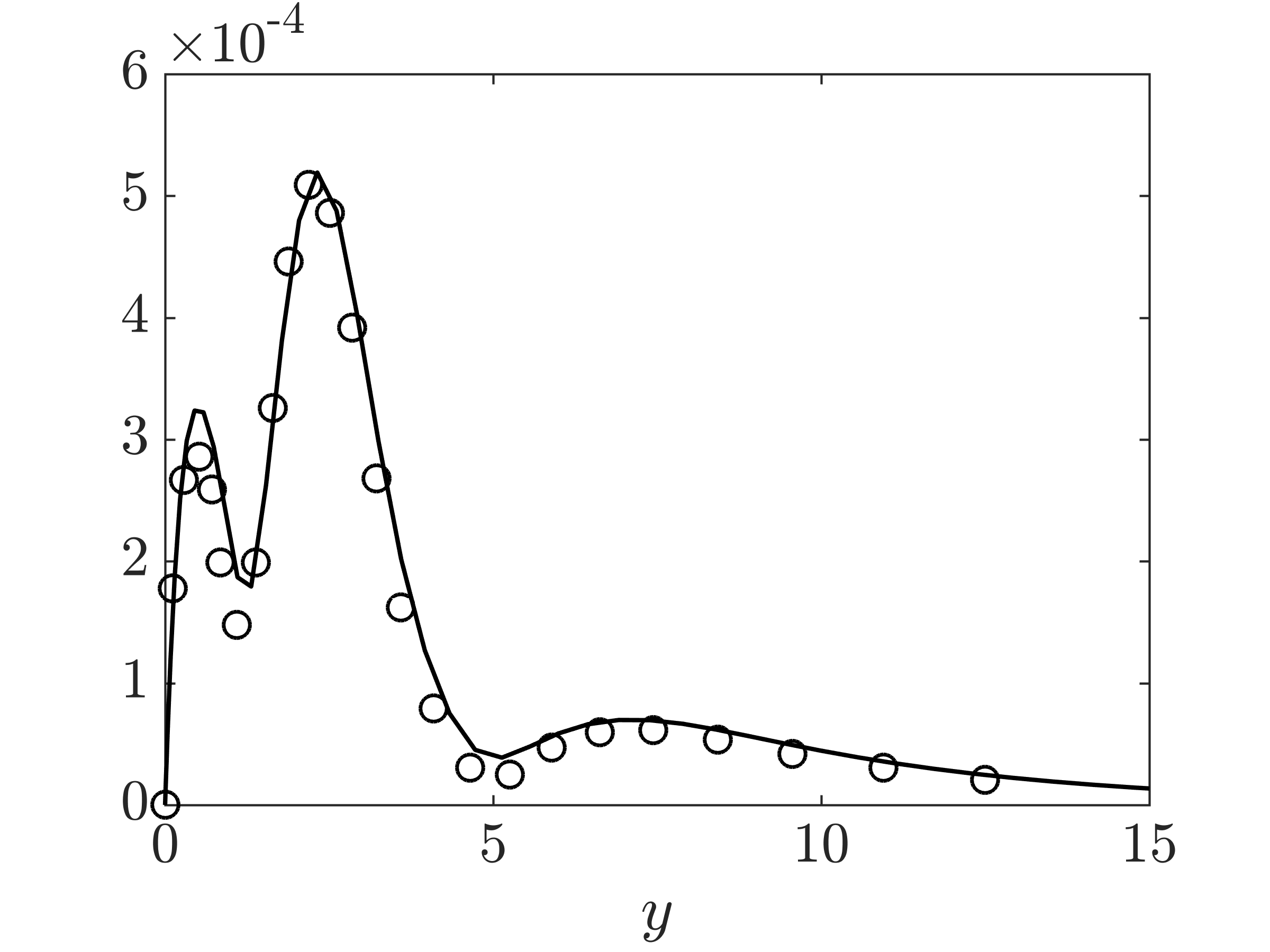}
        \end{tabular}
        \end{tabular}
                \caption{(a) The rms amplitudes of the streamwise velocity components of the fundamental and subharmonic modes resulting from experiments~\cite{kaclev84} ($\vartriangle$), DNS~\cite{sayhammoi13} ($--$), nonlinear PSE~\cite{josstrcha93} ($\circ$), PFE (thick solid lines), and linear PSE (thin solid lines). The fundamental ($2, 0$) mode, ($1, 1$) subharmonic, and ($3, 1$) subharmonic are represented by black, blue, and red colors, respectively. (b-d) The normalized amplitudes of the streamwise velocity components of the ($2,0$) (b), ($1,1$) (c), and ($3,1$) (d) modes at $Re=620$ resulting from PFE ($-$), and nonlinear PSE ($\circ$).}
        \label{fig.om0p0248subharmonic}
\end{figure}

Figure~\ref{fig.om0p0248subharmonicamp} shows the rms amplitude of individual modes resulting from experiments~\cite{kaclev84}, DNS~\cite{sayhammoi13}, nonlinear PSE~\cite{josstrcha93}, along with the present PFE computations. Here, the modes are denoted by ($l,k$), where $l$ stands for the temporal frequency of the harmonic and subharmonic modes as a multiple of the subharmonic mode frequency $\omega=0.248$ and $k$ represents the spanwise wavenumber as a multiple of the fundamental wavenumber $\beta=0.132$. The amplitude of the ($1, 1$) mode from the PFE is in excellent agreement with other results. While the amplitude of the ($3, 1$) mode is somewhat under-predicted, the general trend in the growth of this mode is captured by the PFE. We note that in the absence of interactions between harmonics, the linear PSE results in inaccurate predictions for the amplitude of subharmonic modes; see thin solid lines in Fig.~\ref{fig.om0p0248subharmonicamp}. Figures~\ref{fig.om0p0248subharmonic}(b-d) show the amplitude of the streamwise velocity profile of the modes considered in this study normalized by the results of nonlinear PSE. For all three modes, the profiles resulting from PFE are in good qualitative agreement with the result of nonlinear PSE.

	\vspace*{-2ex}
\section{Streamwise elongated laminar streaks}
\label{sec.Nlstreak}

Bypass transition often originates from non-modal growth mechanisms that can lead to streamwise elongated streaks; see for example~\cite{jocdur01}. The streaks can attain substantial amplitudes (15-20\% of the free-stream velocity) and make the flow susceptible to the amplification of high frequency secondary instabilities~\cite{asakonoiznis07,haczak14}. Secondary instability analysis of saturated streaks has been previously used to analyze the breakdown stage in the transition process~\cite{andbrabothen01,bra-phd03}. However, nonlinear effects that influence the formation of streaks become prominent in earlier stages of transition and before the breakdown of streaks. In this section, we utilize the PFE to capture the interactions between various modes in the amplification of the streaks. We focus on the interaction between different spanwise harmonics and study their contribution to the mean flow distortion (MFD), which in turn affects the energy balance among various harmonics that form streaks. We show that the linear PSE fail to predict such a phenomenon and demonstrate how the PFE provide the means to capture the correct trend in the MFD as well as the resulting velocity distribution.

	\vspace*{-2ex}
\subsection{Setup}
\label{sec.streak-setup}

We trigger the formation of streaks by imposing an initial condition computed via the PSE-based optimization approach introduced in~\cite{hacmoi17}. This optimal initial condition yields the highest amplification of perturbation kinetic energy and it is obtained from the singular value decomposition of a pseudo-propagator which advances arbitrary superpositions of the most unstable eigenfunctions in the non-parallel base flow. The initial perturbation field describes a set of counter-rotating streamwise vortices which give rise to the streaks by means of the lift-up mechanism. We compute the spatial evolution of the perturbation field via the linear PSE and use this solution to augment the Blasius boundary layer base flow profile $\bu_0$ for the subsequent PFE computations as
\begin{align}
\label{eq.streakbase}
    \ba{rcl}
            U(x,y)
            &\;=\;&
            U_B(x,y)
            \;+\;
            U_{S,1}(x,y)\,\mre^{\mri \beta z}
            \;+\;
            U_{S,1}^*(x,y)\,\mre^{-\mri \beta z}
            \\[0.25cm]
            V(x,y)
            &\;=\;&
            V_B(x,y)
            \\[0.25cm]
            W(x,y)
            &\;=\;&
            0,
    \ea
\end{align}
which can be written in the following compact form
\begin{align}
	\label{eq.base-floquet-streak}
	\bar{\bu}(x, y, z)
	~=~
	\ds{\sum_{m\,=\,-1}^{1} \bu_m(x, y)\, \mre^{\mri m\beta z}}.
\end{align}
In the linear PSE computations, the real part of the complex wavenumber $\alpha$ is set to zero in accordance with the nature of streamwise elongated streaks and its imaginary part is initialized with a small number (e.g., $10^{-10}$). Moreover, the exponential growth resulting from the imaginary part of $\alpha$ is absorbed into the amplitude of $U_{S,1}(x,y)$ in Eq.~\eqref{eq.streakbase}.

The velocity field of streamwise elongated streaks is dominated by the growth of the streamwise component while wall-normal and spanwise components experience viscous decay. As a consequence, we disregard the normal and spanwise components of the solution to linear PSE, and only use the streamwise component $U_{S,1}(x,y)$ in~\eqref{eq.streakbase}, which is also in agreement with the structure and amplitude of the initial condition. We represent the state in the PFE using the Fourier expansion
\begin{align}
	{\bq}(x, y, z)
	~=~
	\mre^{\mri \alpha x}
	\ds{\sum_{n \, = \, -\infty}^{\infty}\hat{\bq}_n(x,y)\, \mre^{\mri n \beta z}},
	\label{eq.streak-ansatz}
\end{align}
where $\hat{\bq}_0$ is the MFD, higher-order harmonics in the spanwise direction represent various streaks of wavelength $2\pi/ (n \beta)$, and $\alpha=\mri \alpha_i$ is the uniform streamwise wavenumber over all harmonics. Similar to the procedure in Sec.~\ref{sec.H-type-Setup}, we derive the PFE in the form of Eq.~\eqref{eq.PFE} by substituting the ansatz~\eqref{eq.streak-ansatz} and modulated base flow~\eqref{eq.streakbase} into the linearized NS equations and rearranging the governing equations for each harmonic. For this case study, the operators ${\bf L}_F$ and ${\bf M}_F$ in PFE~\eqref{eq.PFE} are provided in Appendix~\ref{sec.appendix1}.

The procedure for obtaining the results presented in the next subsection can be summarized as follows:
\begin{enumerate}
  \item Compute the initial fluctuation field for maximum streamwise growth using the methodology presented in~\cite{hacmoi17}.
  \item March the initial fluctuation field downstream for the linear evolution of the optimal streak using linear PSE.
  \item Augment the Blasius boundary layer profile with the solution to linear PSE to obtain the modulated base flow $\bar{\bu}$ according to Eq.~\eqref{eq.base-floquet-streak}.
  \item Use the same initial condition as the linear PSE in step 2 and its complex conjugate to initialize $\hat{\bq}_{\pm 1}$ in Eq.~\eqref{eq.streak-ansatz} and initialize other harmonics with zero. Also, set the initial growth rate $\alpha(x_0)$ to a small imaginary number.
  \item Use the PFE to compute the spatial evolution of all harmonic modes around the modulated base flow.
\end{enumerate}

	\vspace*{-2ex}
\subsection{Nonlinear evolution of optimal streaks}
\label{sec.streak-evolve}

Although the initial condition imposed at the first downstream location only contains a single spanwise wavenumber, the appreciable amplitudes of the developing streaks lead to modal interactions that introduce additional harmonics and modulate the mean flow. To investigate these harmonic interactions, we consider truncations of the bi-infinite state $\hat{\bq}$ in the PFE~\eqref{eq.PFE} to $2N+1$ harmonics in $z$, i.e., $n=-N, \cdots, N$. We set $N = 3$ and consider a computational domain with $L_x\times L_y= 2000\times 60$, $N_y=80$ collocation points in the wall-normal direction, and a step-size of $\Delta x=15$; see Appendix~\ref{sec.convergence} for a discussion on wall-normal grid-convergence and about the influence that the number of harmonics has on our results.

The temporal frequency, streamwise and fundamental spanwise wavenumbers are set to $\omega=0$, $\alpha=-10^{-10}\mri$, and $\beta=0.4065$, respectively. Note that the small imaginary-valued wavenumber $\alpha$ corresponds to infinitely long structures in the streamwise direction that saturate after a particular streamwise location. Moreover, $\alpha \neq 0$ maintains a well-conditioned downstream progression for the PFE computations. We initialize the PFE computation at $Re_0=467$ with zero initial conditions for all $\hat{\bq}_n$ with $n \neq \pm 1$. The fundamental harmonic $\hat{\bq}_{\pm 1}$ is initialized with the same initial condition as the primary linear PSE computations and with an rms amplitude of $6.4\times10^{-4}$. Since this case study considers the evolution of disturbances with a slowly varying streamwise wavenumber $\alpha$, we set $\alpha_x=0$ for both the primary linear PSE and the subsequent PFE computations. To verify the predictions of our framework, we also conduct direct numerical simulations of the nonlinear NS equations (with the same initial conditions) using a second-order finite volume code with $2049\times 257 \times 257$ grid points in the streamwise, wall-normal, and spanwise dimensions, respectively.

\begin{figure}
	\begin{centering}
	\begin{tabular}{rc}
	\begin{tabular}{c}
		\vspace{.6cm}
	\end{tabular}
	&
	\begin{tabular}{c}
	\includegraphics[width=8cm]{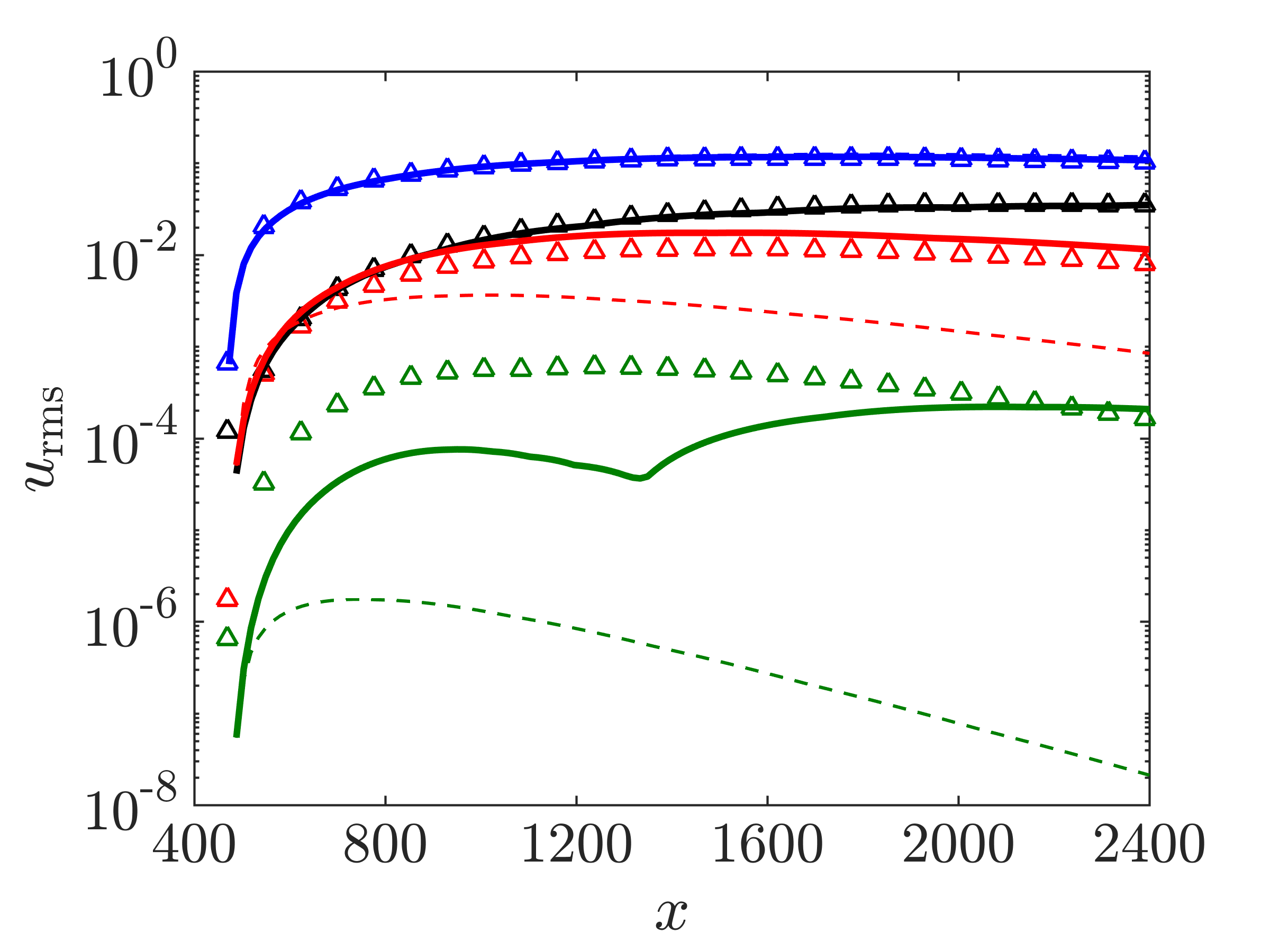}
	\end{tabular}
	\end{tabular}
	\caption{The rms amplitudes of the streamwise velocity components for various harmonics with $\omega=0$ and $\beta=0.4065$ resulting from DNS ($\vartriangle$), PFE ($-$), and linear PSE ($--$). The MFD, first, second, and third harmonics are shown in black, blue, red, and green, respectively.}
	\label{fig.streakamp}
	\end{centering}
\end{figure}

As illustrated in Fig.~\ref{fig.streakamp}, all harmonics undergo an initial algebraic growth followed by saturation. The solution to the linear PSE accurately predicts the evolution of the fundamental spanwise harmonic; cf.~Eq.~\eqref{eq.streak-ansatz}. The PFE accurately predict the growth of the dominant harmonics, and especially the MFD. While a discrepancy is observed for the third harmonic, its contribution to the overall structure of the streaks is negligible. The reasonable prediction of growth trends and generation of the MFD component is a direct consequence of accounting for interactions between different harmonics within our framework because, apart from $\hat{\bq}_{\pm1}$, all other harmonics were initialized with zero.

\begin{figure}
                \begin{tabular}{cccc}
        \subfigure[]{\label{fig.uyzDNS}}
        &&
        \subfigure[]{\label{fig.uyzLPSE}}
        &
        \\[-.2cm]
        \begin{tabular}{c}
                \vspace{.4cm}
        \end{tabular}
        &
        \begin{tabular}{c}
                \includegraphics[width=8cm]{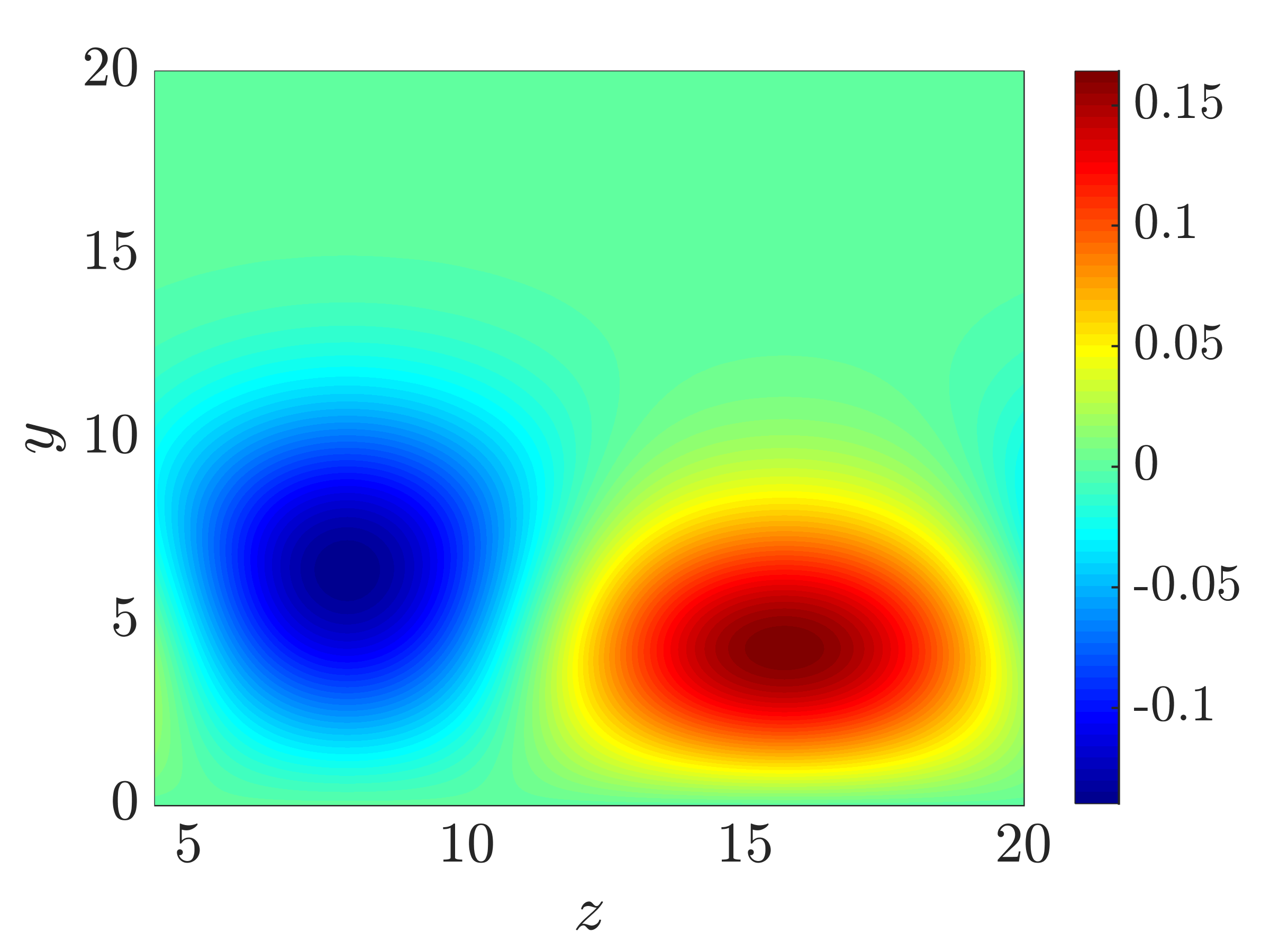}
        \end{tabular}
        &&
        \begin{tabular}{c}
                \includegraphics[width=8cm]{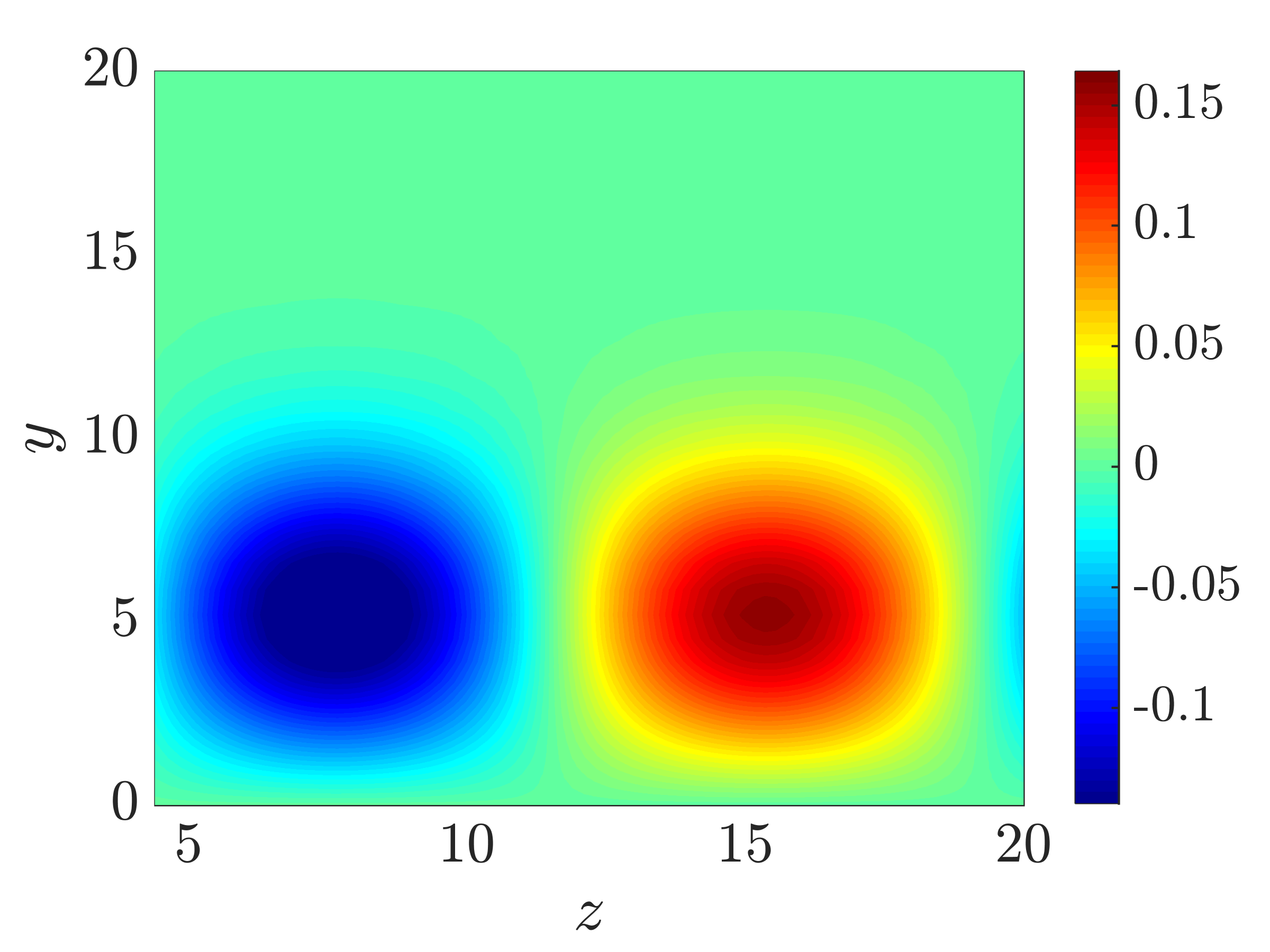}
        \end{tabular}
        \\[-.2cm]
        \subfigure[]{\label{fig.uyzFPSE}}
        &&
        \subfigure[]{\label{fig.uyzFPSEnoMFD}}
        &
        \\[-.2cm]
        \begin{tabular}{c}
                \vspace{.4cm}
        \end{tabular}
        &
        \begin{tabular}{c}
                \includegraphics[width=8cm]{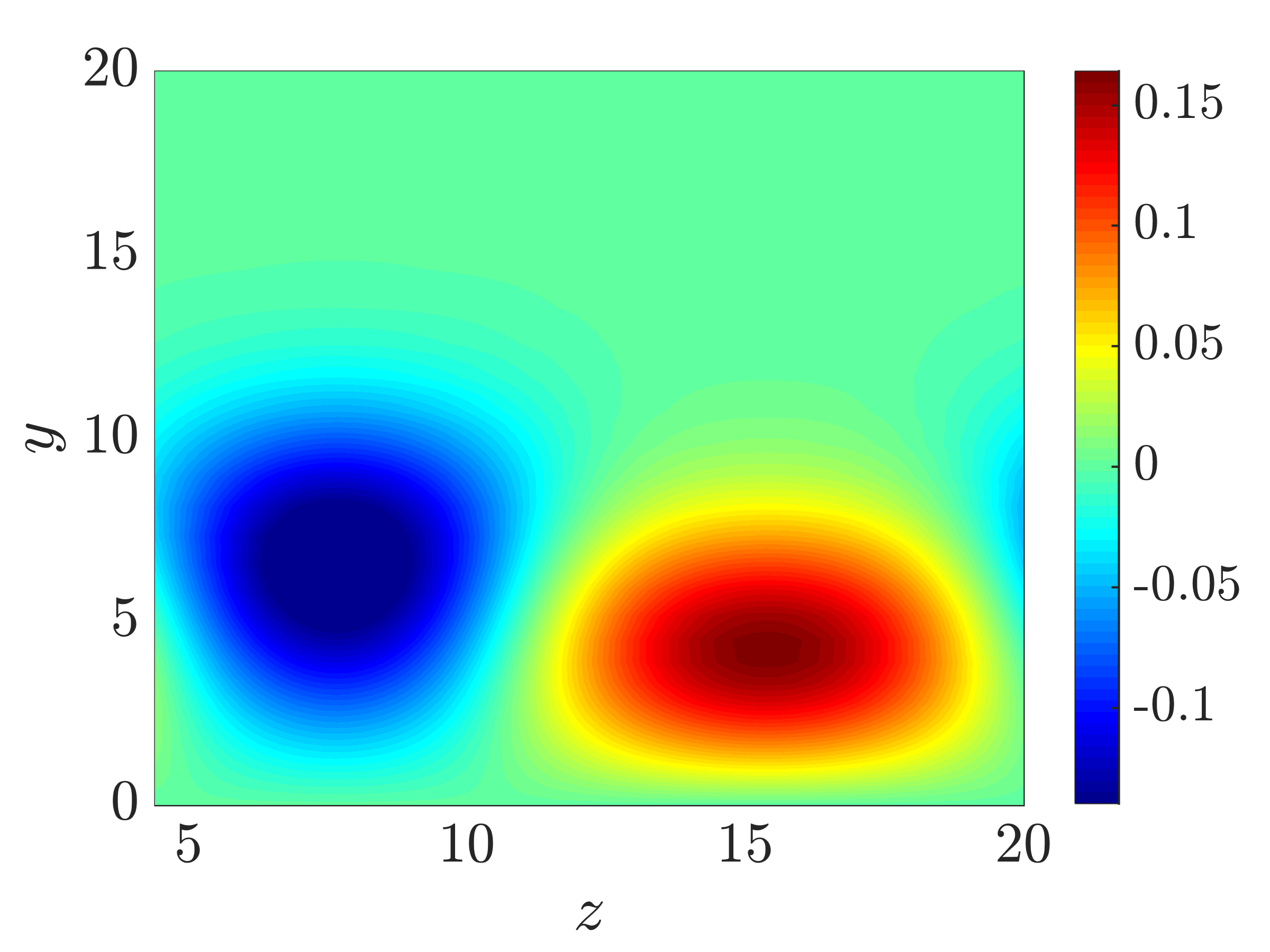}
        \end{tabular}
        &&
        \begin{tabular}{c}
                \includegraphics[width=8cm]{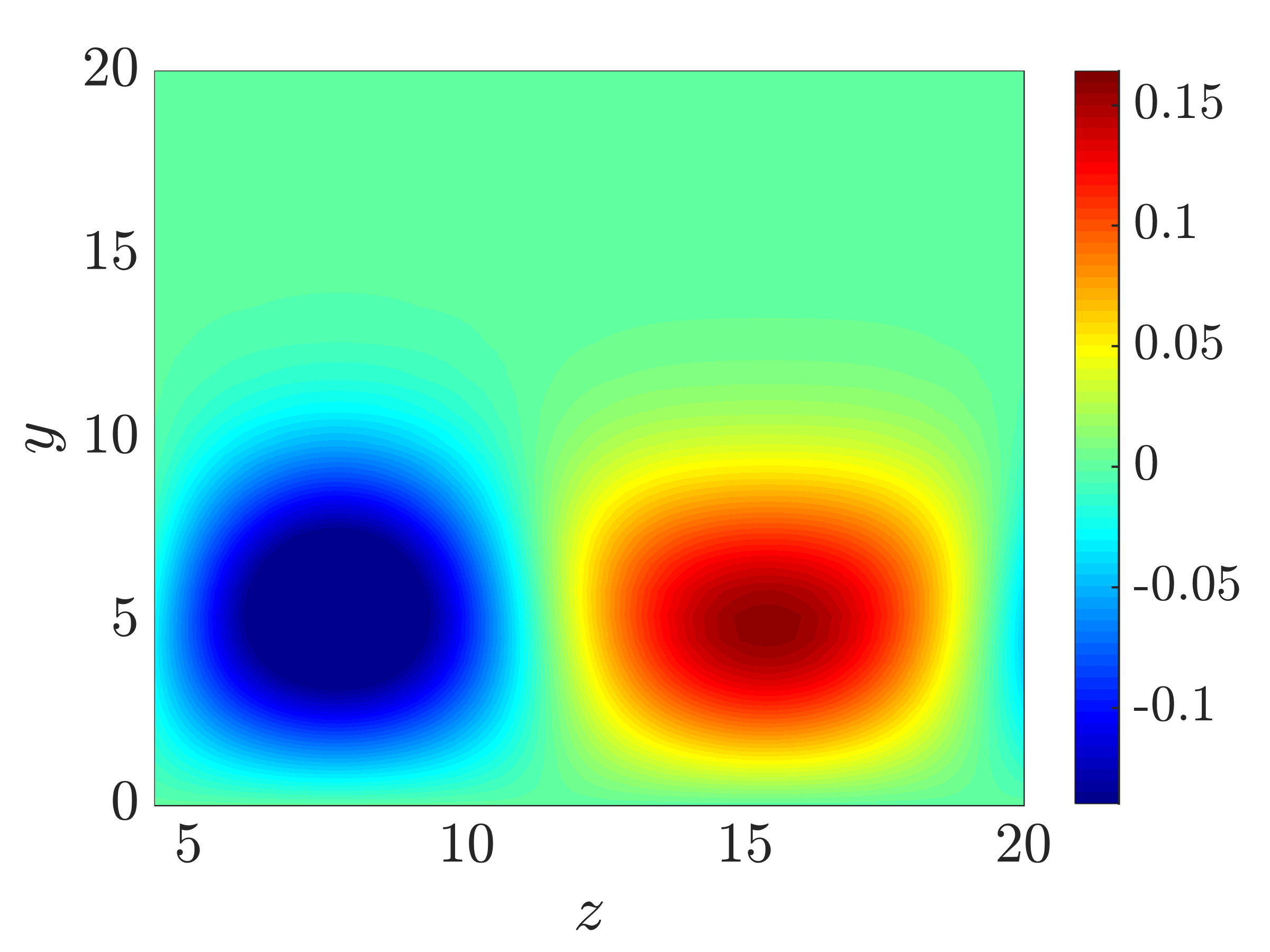}
        \end{tabular}
        \end{tabular}
                \caption{Cross-plane contours of the streamwise velocity of the streaks comprised of all harmonics in the spanwise direction at $x=2400$ resulting from DNS (a), linear PSE (b), and PFE with (c) and without (d) the MFD component.}
        \label{fig.finalstreak}
\end{figure}

Figure~\ref{fig.finalstreak} shows the cross-plane spatial structure of the streaks comprised of all harmonics in the spanwise direction at $x=2400$. Comparison of Figs.~\ref{fig.uyzDNS} and~\ref{fig.uyzLPSE} indicates a significant discrepancy between the shape of the structures in the cross-plane if the interaction between modes is not taken into account. Since the first (fundamental) harmonic has much larger amplitude than the second and third harmonics, the velocity distribution resulting from the linear PSE is dominated by the structure of the first harmonic. Furthermore, in the absence of interactions between harmonics, the linear PSE would not be able to generate the MFD and would thus result in inaccurate predictions for the amplitude of higher-order harmonics; see dashed lines in Fig.~\ref{fig.streakamp}. In Fig.~\ref{fig.streakamp}, we have used a scaled version of the initial condition for the first harmonic to initialize the linear PSE computations for higher-order harmonics. Figure~\ref{fig.uyzFPSE} demonstrates excellent agreement of the results from PFE and DNS. To emphasize the contribution of the MFD on the final velocity distribution, Fig.~\ref{fig.uyzFPSEnoMFD} shows the cross-plane contours of the streamwise velocity component without the MFD. While the influence of the MFD on the growth of the first, second, and third harmonics has been retained, this figure demonstrates its influence on the shape of the streamwise elongated structures.

\begin{figure}
        \begin{tabular}{cccc}
        \subfigure[]{\label{fig.2x123runcompare}}
        &&
        \subfigure[]{\label{fig.streakamp2x}}
        &
        \\[-.2cm]
        \begin{tabular}{c}
                \vspace{.4cm}
        \end{tabular}
        &
        \begin{tabular}{c}
                \includegraphics[width=8cm]{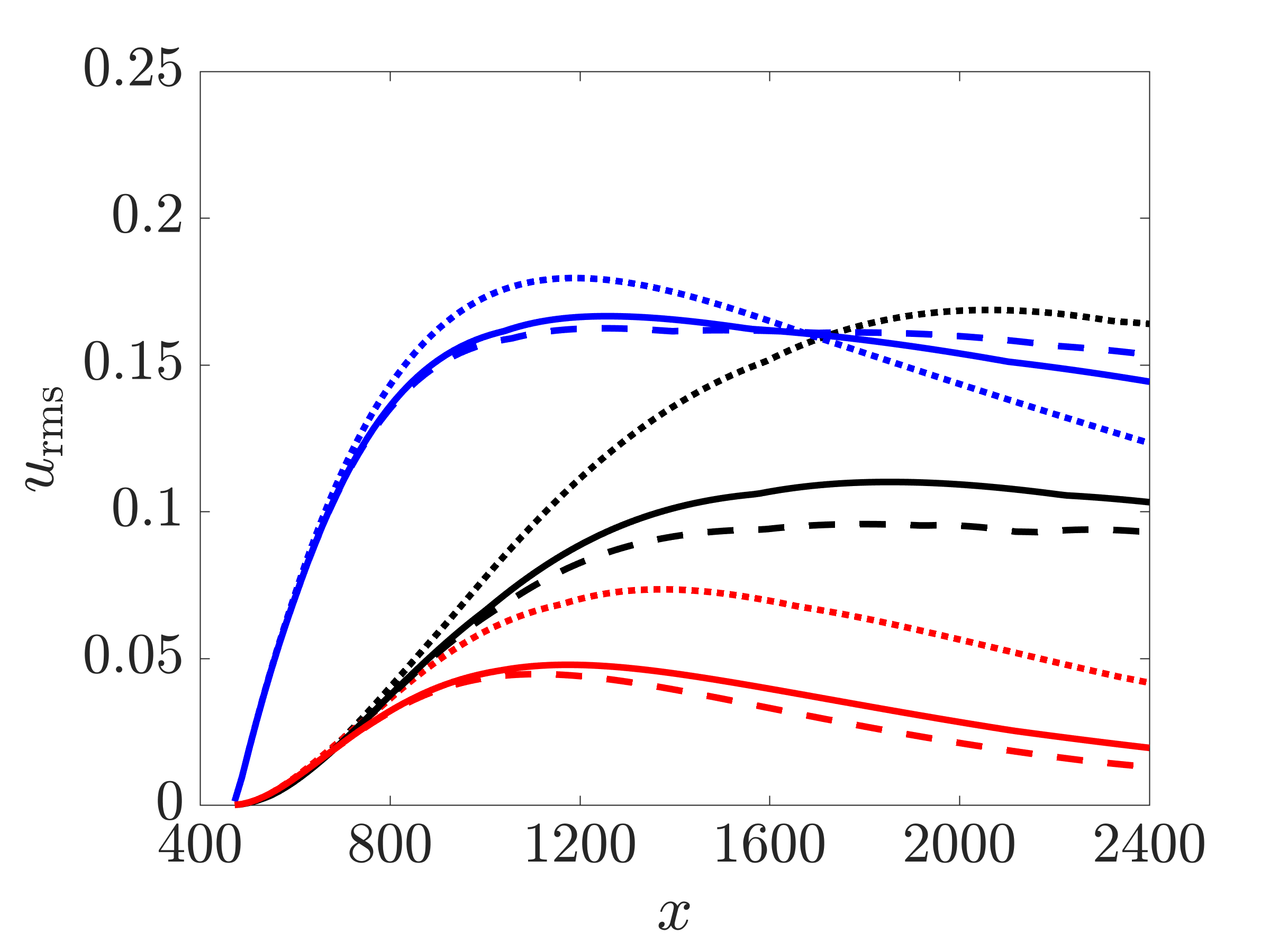}
        \end{tabular}
        &&
        \begin{tabular}{c}
                \includegraphics[width=8cm]{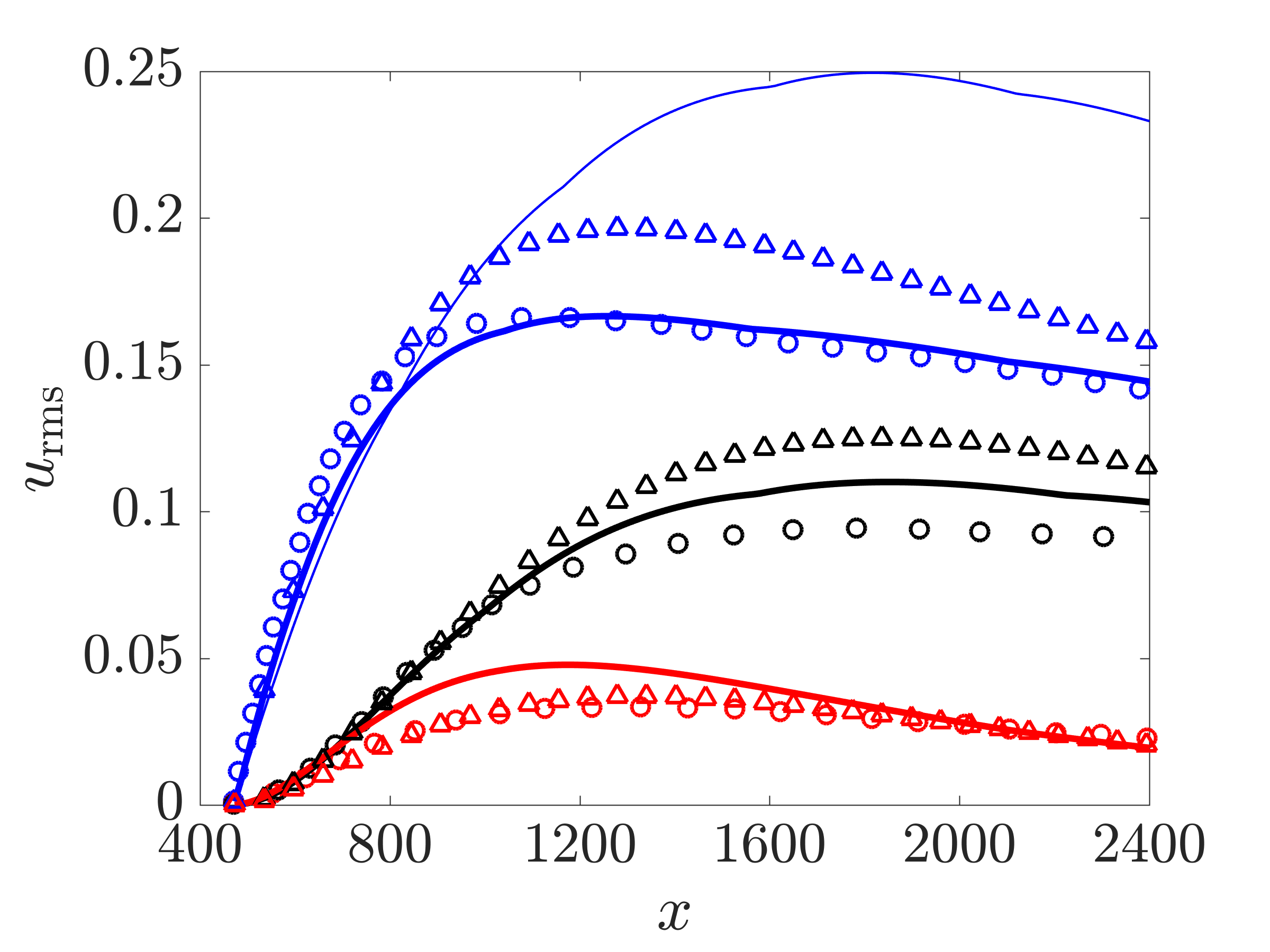}
        \end{tabular}
        \end{tabular}
                \caption{(a) The rms amplitudes of the streamwise velocity components for various harmonics with $\omega=0$ and $\beta=0.4065$ after  {the first ($\cdots$), third ($--$), and seventh ($-$)} run of the PFE. The MFD, first, and second harmonics are shown in black, blue, and red, respectively. (b) The rms amplitudes resulting from DNS ($\vartriangle$), nonlinear PSE ($\circ$), and PFE ($-$). The evolution of the fundamental harmonic due to linear PSE is shown by the thin solid line.}
        \label{fig.finalstreakamp2x}
\end{figure}

Since the evolution of streaks is highly influenced by nonlinear interactions, it is worth examining if the results would change with an increase in the streak amplitude. To test the robustness of our framework we consider a different initial condition which has twice the amplitude as the previous case. We use the same computational configuration as before and initialize all harmonics apart from $\hat{\bq}_{\pm 1}$ with zero. As shown in Fig.~\ref{fig.streakamp2x}, the linear PSE provide a poor prediction for the amplitude of the fundamental harmonic. We use the MFD $\hat{\bq}_{0}$, fundamental harmonics $\hat{\bq}_{\pm 1}$, and second harmonics $\hat{\bq}_{\pm 2}$ from each run of PFE to update the base flow modulation and rerun the PFE (cf. Fig.~\ref{fig.equilibrium}). The base flow in subsequent iterations is thus given by
\begin{align}
\label{eq.streakbase-rerun}
    \ba{rcl}
            U(x,y)
            & \!\! = \!\! &
            U_B(x,y)
            \, + \,
            U_{S,0}(x,y)
            \, + \,
            U_{S,1}(x,y)\,\mre^{\mri \beta z} \, + \,U_{S,1}^*(x,y)\,\mre^{-\mri \beta z}
            \, + \,
            U_{S,2}(x,y)\,\mre^{2\mri \beta z} \, + \,U_{S,2}^*(x,y)\,\mre^{-2\mri \beta z},
            \\[0.25cm]
            V(x,y)
            & \!\! = \!\! &
            V_B(x,y) \, + \, V_{S,0}(x,y),
            \\[0.25cm]
            W(x,y)
            & \!\! = \!\! &
            0,
    \ea
\end{align}
and takes the compact form
\be
	\label{eq.base-floquet-streak-rerun}
	\bar{\bu}(x, y, z)
	~=~
	\ds{\sum_{m\,=\,-2}^{2} \bu_m(x, y)\, \mre^{\mri m\beta z}}.
\ee
Here, $U_{S,0}$, $U_{S,1}$, and $U_{S,2}$ represent the MFD, first- and second-order harmonics corresponding to the solution of the previous PFE run, respectively, and $\bu_0$ contains both the Blasius profile and the MFD. We note that higher-order harmonics are omitted from the base flow modulation in Eq.~\eqref{eq.base-floquet-streak-rerun} as they do not significantly influence the profiles that result from the iterative PFE procedure. Furthermore, similar to Eq.~\eqref{eq.streakbase}, special care is taken in modulating the base flow, i.e., higher-order harmonics ($|m|\geq 1$) are excluded from the wall-normal and spanwise components of the base flow modulation in agreement with the structure and amplitude of the initial condition.

Figure~\ref{fig.2x123runcompare} demonstrates how iterating the PFE can improve our prediction of the predominantly nonlinear streak evolution. The rms curves for various harmonics converge after $7$ iterations of the PFE feedback loop illustrated in Fig.~\ref{fig.equilibrium}. Note that in the previous case of moderate-amplitude streaks, subsequent iterations were not necessary and accurate results were obtained after one run of the PFE over the streamwise domain. The high number of iterations required for convergence is indicative of the significant role nonlinear terms play in the more challenging case of high-amplitude streaks. In Fig.~\ref{fig.streakamp2x}, we compare the result from the final iteration (solid lines in Fig.~\ref{fig.2x123runcompare}) with the result of DNS and nonlinear PSE. We see that the PFE capture the initial algebraic growth, inhibition of growth, and general trend in the saturation of amplitudes.

\begin{figure}[!ht]
        \begin{tabular}{cccc}
        \subfigure[]{\label{fig.uyzDNS2x}}
        &&
        \subfigure[]{\label{fig.uyzFPSE2x}}
        &
        \\[-.2cm]
        \begin{tabular}{c}
                \vspace{.4cm}
        \end{tabular}
        &
        \begin{tabular}{c}
                \includegraphics[width=8cm]{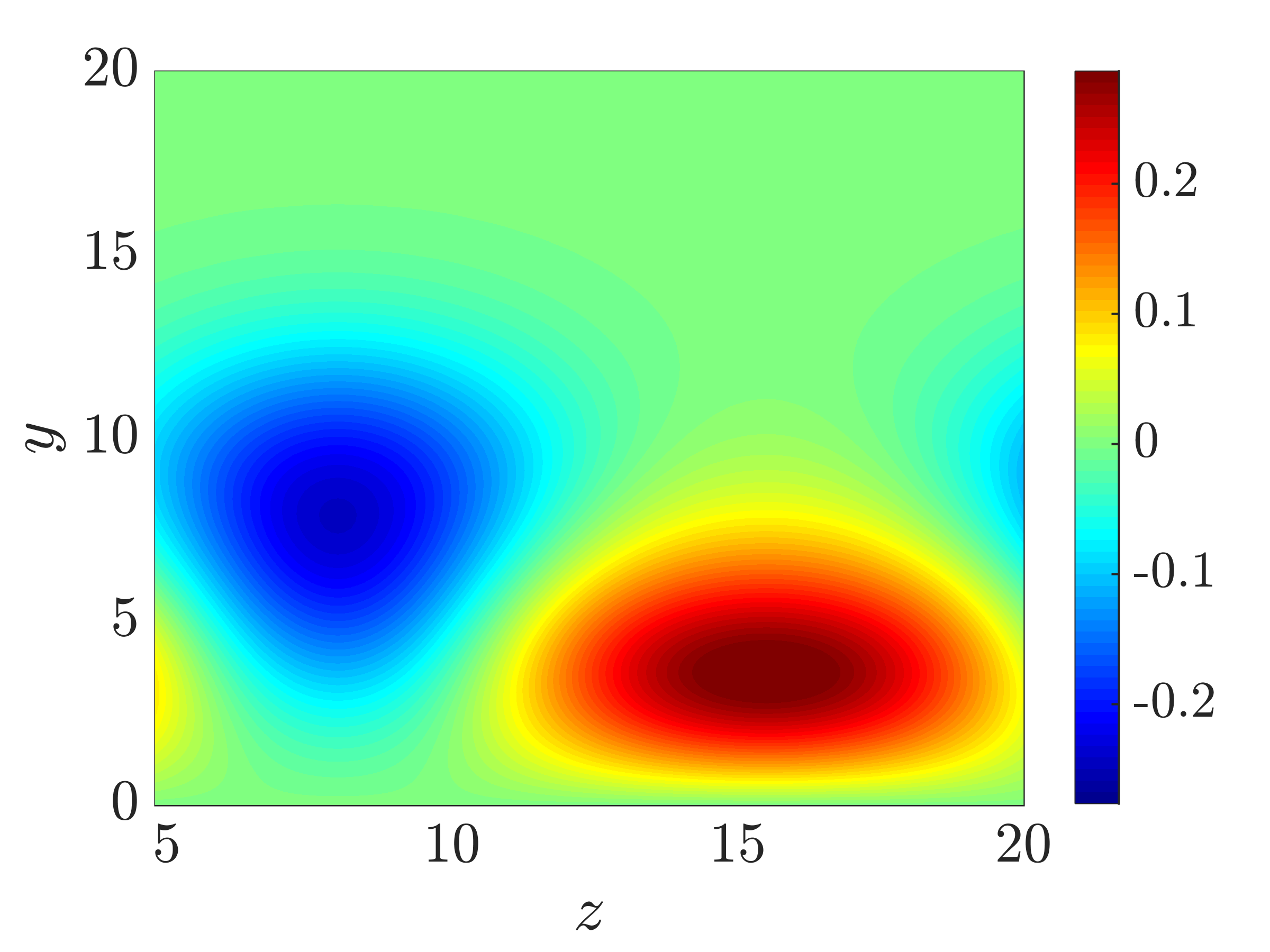}
        \end{tabular}
        &&
        \begin{tabular}{c}
                \includegraphics[width=8cm]{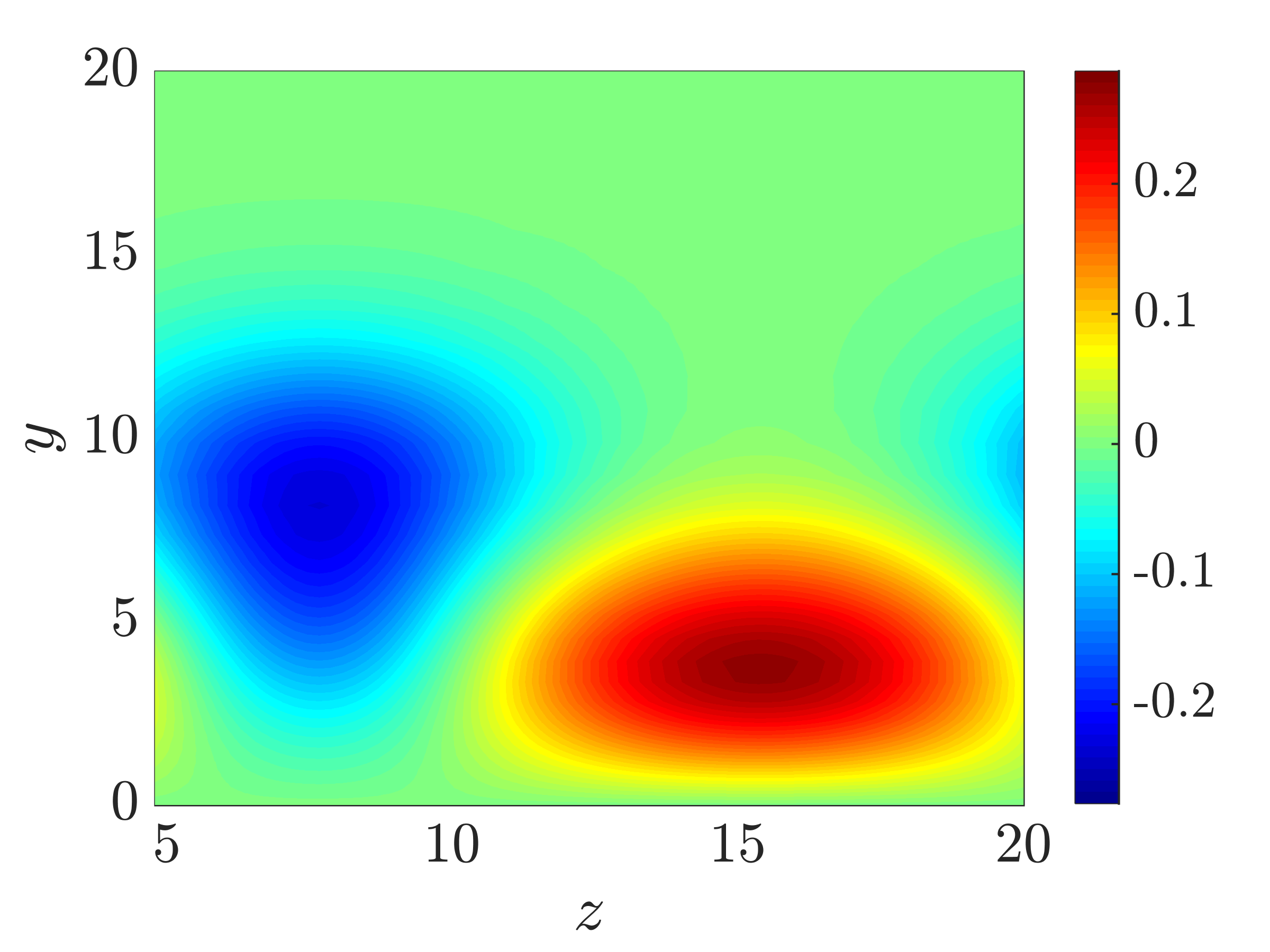}
        \end{tabular}
        \end{tabular}
                \caption{Cross-plane contours of the streamwise velocity of the higher amplitude streaks at $x=2400$, which is comprised of all harmonics in the spanwise direction; (a) DNS and (b) PFE.}
        \label{fig.finalstreak2x}
\end{figure}

\begin{figure}[!ht]
        \begin{tabular}{cccc}
        \subfigure[]{\label{fig.2xumode0id81}}
        &&
        \subfigure[]{\label{fig.2xumode1id81}}
        &
        \\[-.2cm]
        \begin{tabular}{c}
                \vspace{.4cm}
        \end{tabular}
        &
        \begin{tabular}{c}
                \includegraphics[width=8cm]{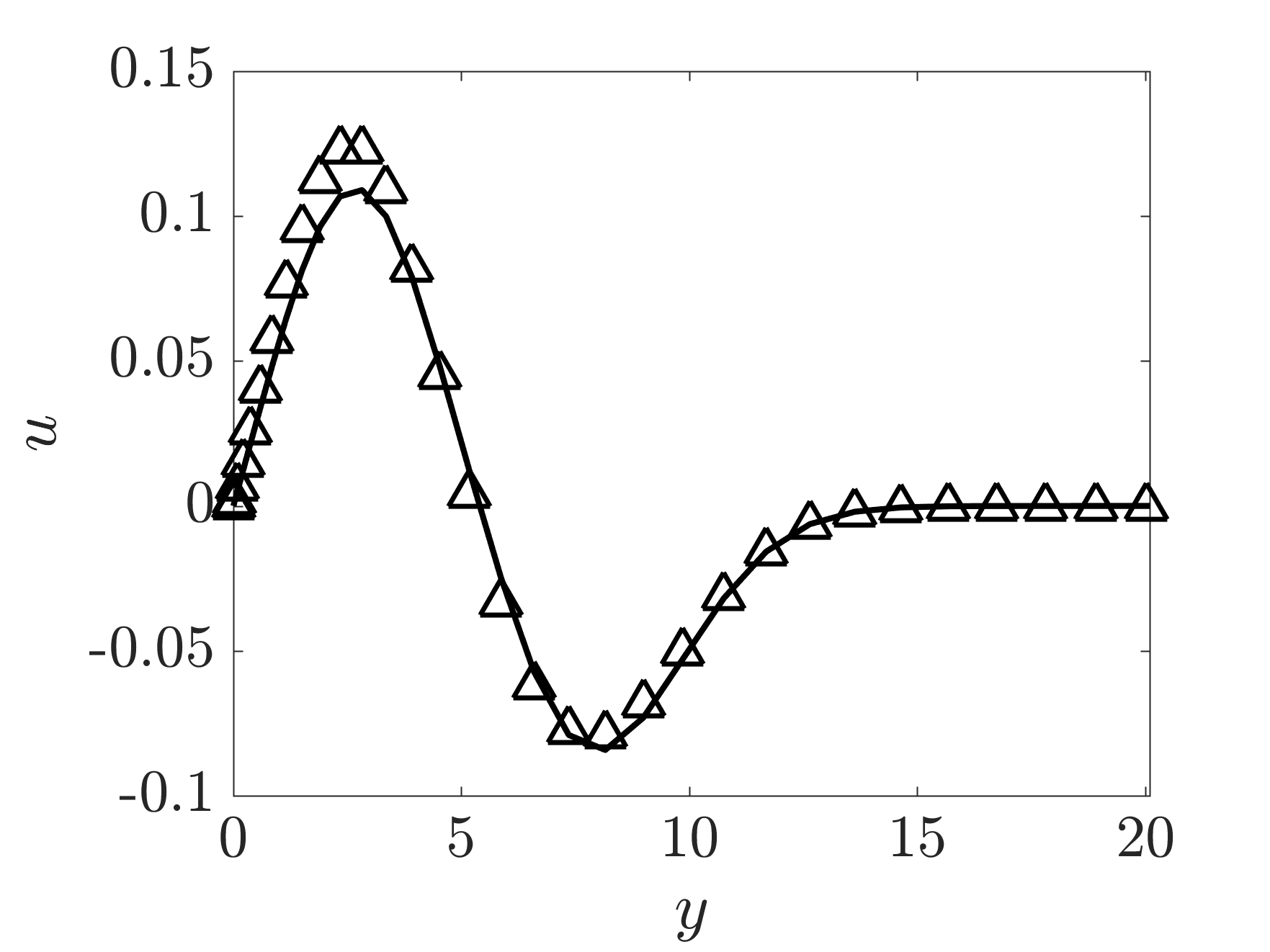}
        \end{tabular}
        &
        \begin{tabular}{c}
                \vspace{.4cm}
        \end{tabular}
        &
        \begin{tabular}{c}
                \includegraphics[width=8cm]{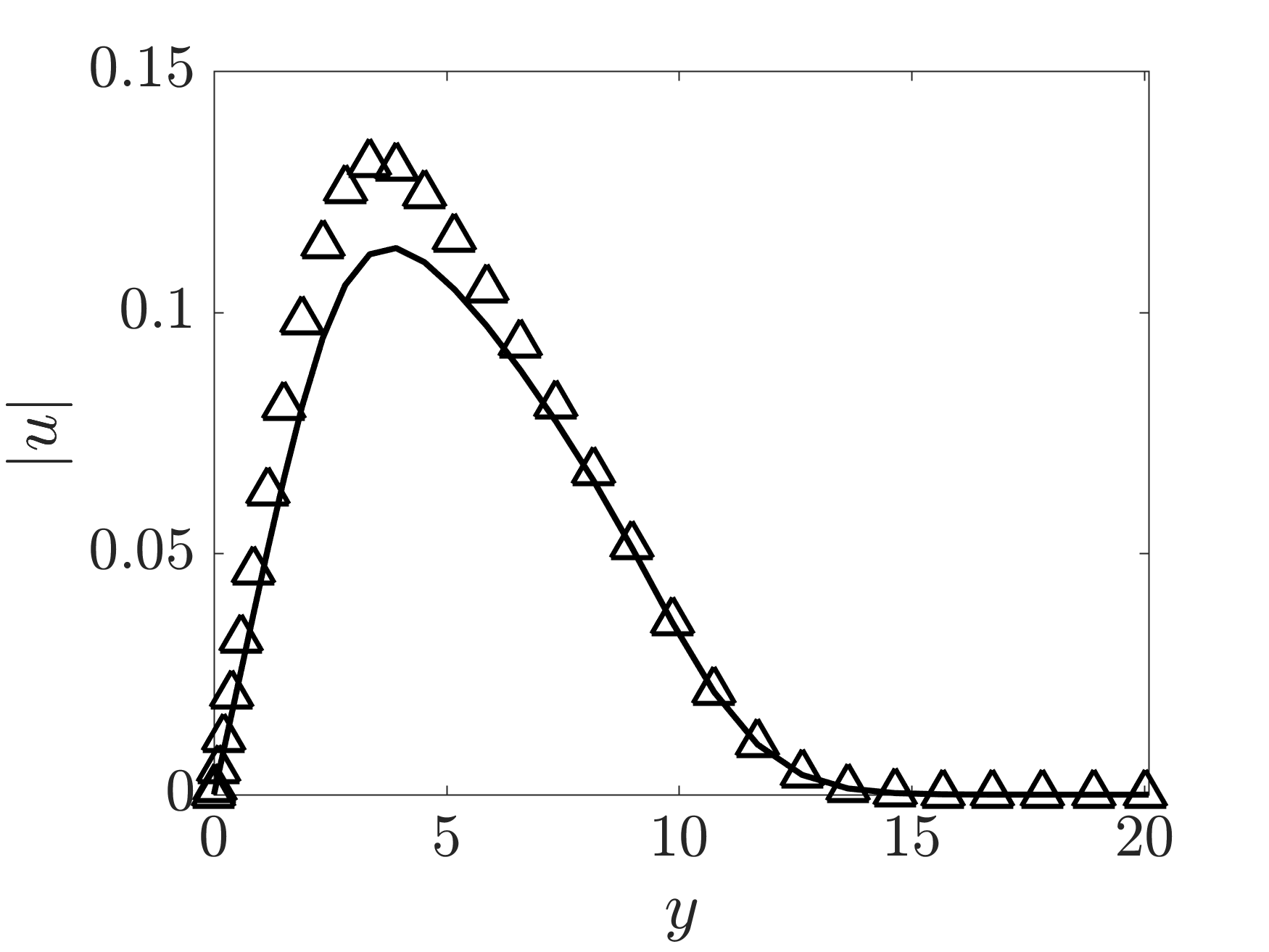}
        \end{tabular}
        \end{tabular}
        \caption{ {(a) The streamwise velocity component of the MFD; and (b) the magnitude of the streamwise component of the first harmonic at $x=1700$ resulting from DNS ($\vartriangle$) and PFE ($-$).}}
        \label{fig.streak2xprofileid81}
\end{figure}

\begin{figure}[!ht]
        \begin{tabular}{cccc}
        \subfigure[]{\label{fig.2xumode0id129}}
        &&
        \subfigure[]{\label{fig.2xumode1id129}}
        &
        \\[-.2cm]
        \begin{tabular}{c}
                \vspace{.4cm}
        \end{tabular}
        &
        \begin{tabular}{c}
                \includegraphics[width=8cm]{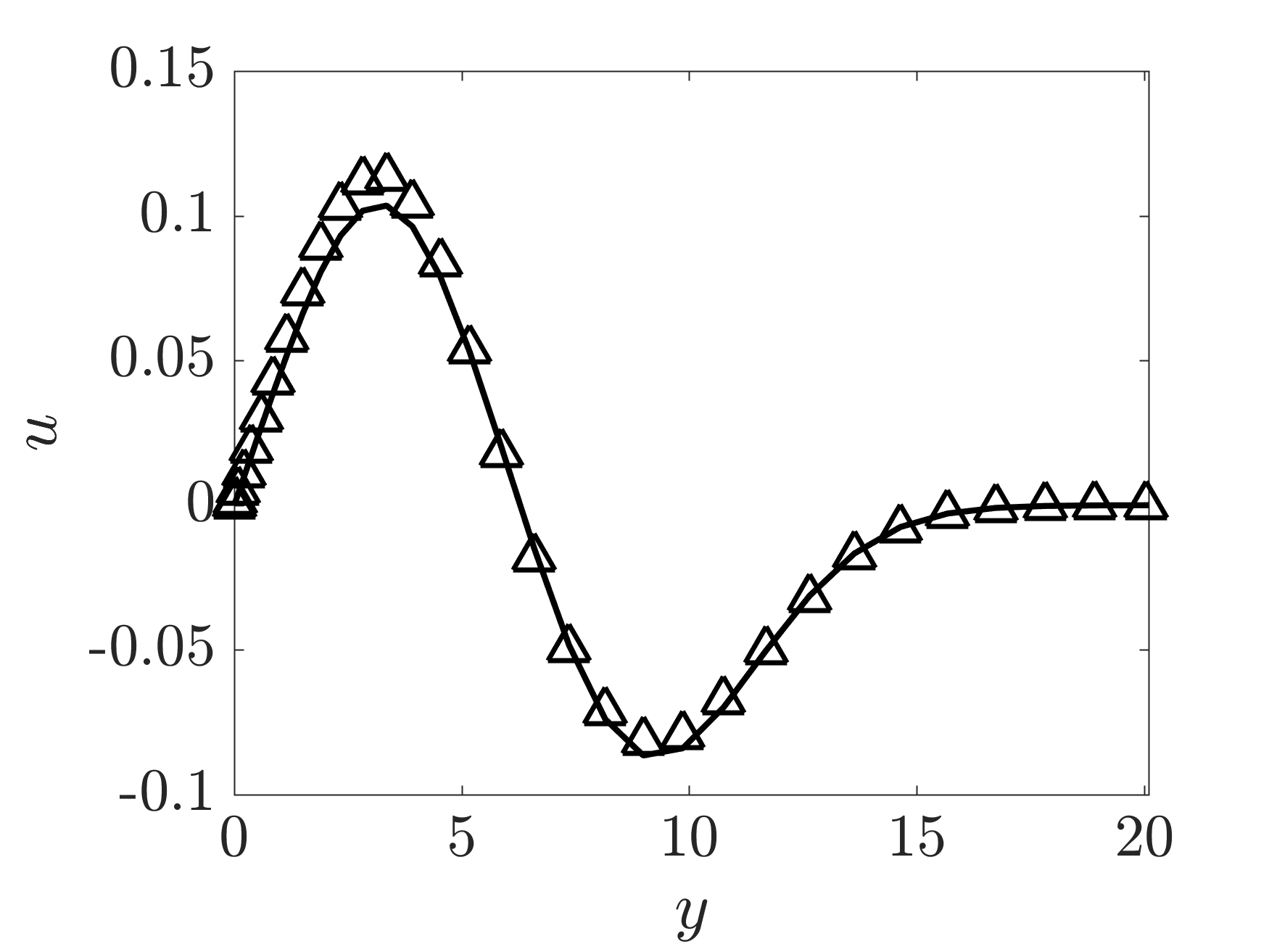}
        \end{tabular}
        &
        \begin{tabular}{c}
                \vspace{.4cm}
        \end{tabular}
        &
        \begin{tabular}{c}
                \includegraphics[width=8cm]{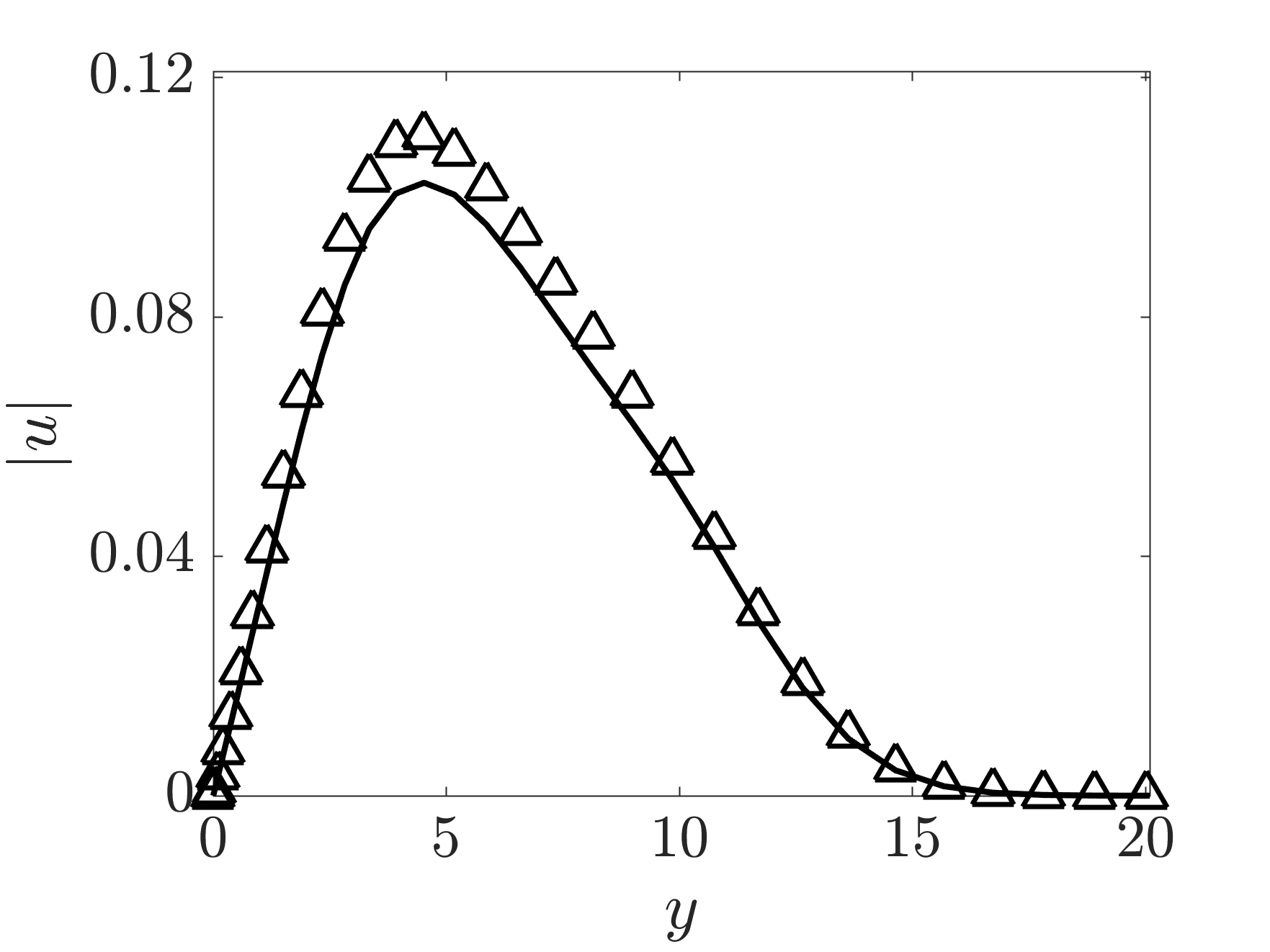}
        \end{tabular}
        \end{tabular}
        \caption{(a) The streamwise velocity component of the MFD; and (b) the magnitude of the streamwise component of the first harmonic at $x=2400$ resulting from DNS ($\vartriangle$) and PFE ($-$).}
        \label{fig.streak2xprofileid129}
\end{figure}

Nonlinear interactions generate an appreciable MFD that alters the mean flow profile and hampers the growth of the principal harmonic in comparison to the single mode computation of linear PSE (cf.\ thin blue line in Fig.~\ref{fig.2x123runcompare}). Previous studies have reported the stabilizing effect of nonlinearity on the evolution of unsteady streaks~\cite{ricluowu11} and the boundary layer response to perturbations~\cite{leiwungol99b,zucbotluc06}. In Fig.~\ref{fig.streakamp2x}, while a discrepancy is observed in the prediction of the MFD, dominant trends in the shape and amplitude of velocity profiles are consistently captured in the streamwise domain, and the final amplitudes are in close agreement with the result of nonlinear PSE. However, both nonlinear PSE and PFE seem to under-predict the growth of the fundamental harmonic and MFD. Figures~\ref{fig.streak2xprofileid81} and~\ref{fig.streak2xprofileid129} show the streamwise velocity component of the MFD and first harmonic at $x=1700$ and $x=2400$, which correspond to the largest error in matching the MFD and the end of the longitudinal domain. Finally, as the cross-plane contour plots of Fig.~\ref{fig.finalstreak2x} demonstrate, the PFE provide good predictions for the spatial structure of the streaks that are comprised of various harmonics.

In the present case study, nonlinear interactions play a crucial role in the growth of high amplitude streaks. The PFE capture the nonlinear interactions by allowing spanwise modulations to the base state and extending the state variable $\hat{\bq}$ over multiple harmonics in the spanwise direction. Subsequent iterations of the PFE feedback loop (Fig.~\ref{fig.equilibrium}) refine our predictions of nonlinear interactions on a sweep-by-sweep basis, i.e., by treating the base flow as a streamwise varying parameter in each individual PFE run, and only updating it for the next run. This is in contrast to nonlinear PSE in which nonlinear interactions are captured by explicitly converging over the corresponding nonlinear terms at each step of the streamwise progression. Regardless of how nonlinear interactions are captured, our results demonstrate the difficulty in accurately capturing the correct growth of these optimal streaks (cf. Fig.~\ref{fig.streakamp2x}). While the approximation used by the PFE framework may be seen as a limitation, the encouraging performance of the PFE warrants future study into improving the predictive capability of models that capture harmonic interactions through iterative refinement of the base state and not by explicitly computation of nonlinear terms.

\vspace*{-2ex}
\section{Comparison with nonlinear PSE}
\label{sec.comparison-npse}

In contrast to the nonlinear PSE, which treat the interaction between various modes as a forcing, the PFE introduced in~\eqref{eq.PFE} account for a subset of dominant interactions between the primary and secondary modes while maintaining the linear progression of the governing equations. Compared to the nonlinear PSE, the implementation and evaluation of the PFE is thus less complex as the explicit evaluation of the nonlinear terms and the commonly used transforms between physical and Fourier domains are avoided. More specifically, at each downstream location, the PSE can be viewed as a predictor-corrector algorithm that iterates over nonlinear terms. As the amplitudes of the harmonics grow, these iterations may fail to converge and the PFE does not rely on them. There have been previous efforts to suppress the feedback from secondary to primary modes and to maintain the march of nonlinear PSE through the transitional region; see for example~\cite[Sec.~3.4.3]{her94}. The framework advanced in the current paper allows for the formal investigation of such effects by limiting the interactions within the PFE framework to a subset of dominant harmonics of the base flow. While in practice we observe that the PFE alleviate challenges that may arise from nonlinear interactions, a rigorous proof of convergence for the PFE iterations is deferred to future research.

{The computational cost of the PSE is dominated by the inner iterations that are required to evaluate nonlinear terms at each step of the marching procedure. In contrast, the PFE are advanced by inverting a sparse matrix of higher dimension at each step; the worst-case complexity analysis for dense matrices would suggest that the PFE need more operations per iteration. However, our computational experiments show that even without exploiting the sparse structure of the matrices, our PFE computations require approximately the same amount of time to converge as nonlinear PSE computations. Further improvement of the computational efficiency of our method in a way that would lead to a fair comparison to the PSE is out of the scope of the current~work.}

	\vspace*{-2ex}
\section{Concluding remarks}
\label{sec.conclusion}

We have combined ideas from Floquet decomposition and the linear PSE to develop the parabolized Floquet equations which can be used to march primary and secondary instability modes while accounting for dominant mode interactions. Our modeling framework involves two steps: (i) the linear PSE are used to march the primary disturbances in the streamwise direction; (ii) the PFE are used to march velocity fluctuations around the modulated base flow profile while capturing weakly nonlinear effects and the interaction of modes. The developed framework can account for secondary instabilities as fluctuations around a modulated base flow that includes primary modes generated in step (i). {The PFE involve a linear march of various harmonics} and can be used as a tool to decipher the role of individual harmonics in the spatial evolution of the fluctuation field. Furthermore, subsequent iterations of the PFE, in which the base flow modulation results from the previous PFE computation, can provide a corrective sequence that improve the quality of prediction. To demonstrate the utility of the proposed modeling framework, we have examined the secondary instability analysis of the H-type transition scenario and the evolution of streamwise streaks. Our computational experiments demonstrate good agreement with DNS and nonlinear PSE.

We note that the overall performance of the proposed method relies on the reasonable prediction of the evolution of primary disturbances using linear PSE. In cases where the linear PSE give a poor prediction, an additional source of white or colored stochastic excitation can be used to replicate the effect of nonlinearities and improve the outcome of linear PSE; see~\cite[Sec.~IV]{ranzarhacjovACC17}. For this purpose, the spatio-temporal spectrum of stochastic excitation sources can be identified using the recently developed theoretical framework outlined in~\cite{zarchejovgeoTAC17,zarjovgeoJFM17}.
This methodology can also be used to improve the accuracy of results when nonlinear interactions are of critical importance in the evolution of multimodal dynamics. Implementation of such ideas to further improve the predictive capability of the proposed method is a topic for future research.

\begin{figure}
\begin{centering}
\includegraphics[height=4.3cm]{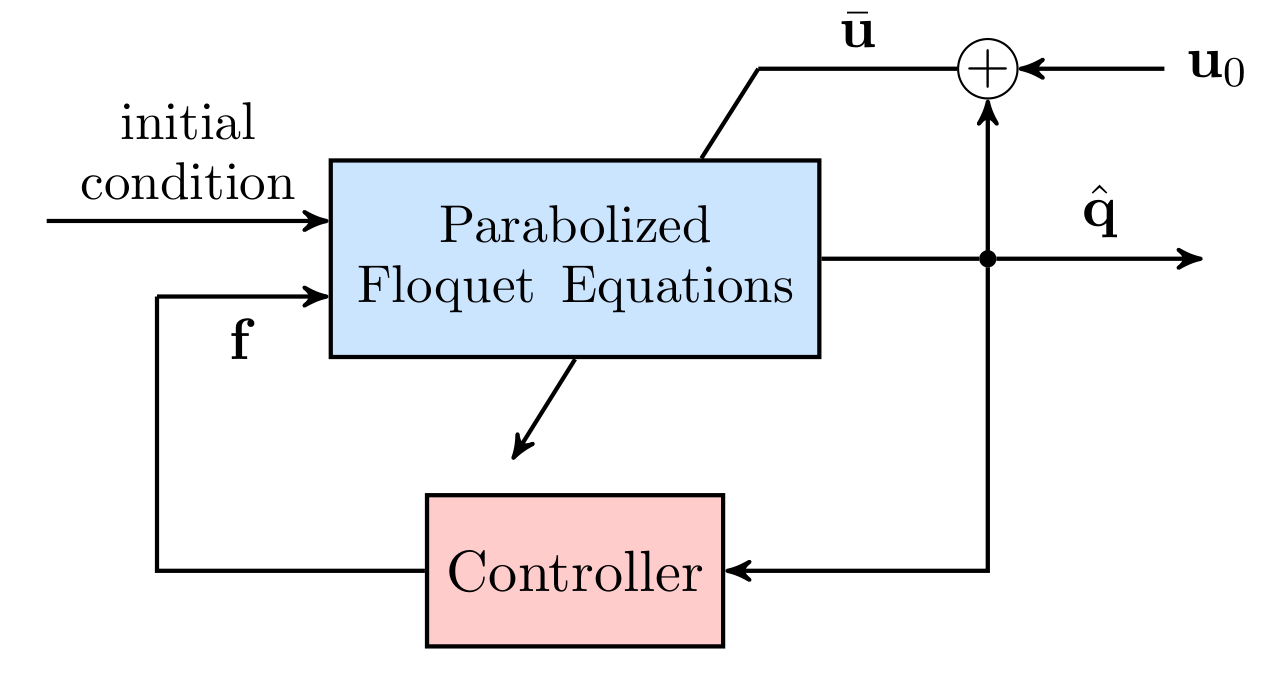}
\caption{Block diagram illustrating the inclusion of control into the PFE loop. The controller uses the fluctuation field resulting from previous PFE iterations to dictate the control signal ${\bf f}$ which perturbs the flow dynamics.}
	\label{fig.PFEdiagram-equilibrium-control}
\end{centering}
\end{figure}

In the PFE framework, nonlinear interactions are captured via the interplay between the state and periodic base flow. The equilibrium configuration in Fig.~\ref{fig.equilibrium} provides the means to better approximate nonlinear interactions through iterative refinement of the base state. While this configuration implies that the PFE framework is inherently nonlinear, each run of the PFE is linear and is thus well-suited for feedback control design using the tools from linear systems theory. More specifically, the modes that are marched using PFE modulate the base state as a streamwise varying parameter in subsequent PFE runs and thus should not be assumed as variables that violate the premise of linearity. Based on the fluctuation field generated at each sweep of PFE, an optimal control strategy can be synthesized to perturb the dynamics of subsequent PFE runs; see schematic in Fig.~\ref{fig.PFEdiagram-equilibrium-control}. While both the control strategy and dynamics are simultaneously updated, convergence of the fluctuation field $\hat{\bq}$ would insure that the final control design is indeed optimal. Analyzing the performance of this design strategy and providing theoretical justification for convergence calls for additional in-depth examination.

\begin{acknowledgments}
Part of this work was conducted during the 2016 CTR Summer Program with financial support from Stanford University and NASA Ames Research Center. We thank Prof.\ P.\ Moin for providing us with the opportunity to participate in the CTR Summer Program and Prof.\ P.\ J.\ Schmid and Dr.\ A.\ Towne for insightful discussions. Financial support from the National Science Foundation under Award CMMI 1739243 and the Air Force Office of Scientific Research under Awards FA9550-16-1-0009 and FA9550-18-1-0422 is gratefully acknowledged.
\end{acknowledgments}

	\vspace*{-2ex}
\appendix

\section{Operators for H-type transition}
\label{sec.appendix2}

For the case study considered in Sec.~\ref{sec.H-type-transition}, the operators ${\bf{L}}_{n,m}$ and ${\bf{M}}_{n,m}$ in ${\bf{L}}_F$ and ${\bf{M}}_F$ from Eq.~\eqref{eq.PFE} are of the form:
\begin{subequations}
	\begin{eqnarray}
	\non
	{\bf L}_{n,0}
	&=&
        \left[
                \ba{cccc}
                \Gamma_n \,-\, \partial_{y}V_B & \partial_{y}\, U_B & 0 & \mri \left((n+0.5)\alpha_r \,+\, \gamma \right)
                \\[.2cm]
                0 & \Gamma_n \,+\, \partial_{y}V_B & 0 & \partial_{y}
                \\[.2cm]
                0 & 0 & \Gamma_n & \mri \beta
                \\[.2cm]
                \mri \left((n+0.5)\alpha_r \,+\, \gamma \right) & \partial_{y} & \mri \beta & 0
                \ea
        \right],
	\\[.35cm]
	\non
	{\bf L}_{n,-1}
	&=&
	\left[
        \ba{cccc}
        \Gamma_{n-} - \partial_{y} V_T \,&\, \partial_{y}\,U_T & 0 & 0
        \\[.2cm]
        \mri\alpha_r V_T \,&\, \Gamma_{n-}  + \partial_{y}V_T  & 0 & 0
        \\[.2cm]
        0 & 0 & \Gamma_{n-}  & 0
        \\[.2cm]
        0 & 0 & 0 & 0
        \ea
        \right],
        \;
        {\bf L}_{n,+1}
        \,=
	\left[
	\ba{cccc}
                \Gamma_{n+} - \partial_{y} V_T^* \,&\, \partial_{y}\,U_T^* & 0 & 0
                \\[.2cm]
                -\mri\alpha_r V_T^* \,&\, \Gamma_{n+}  + \partial_{y}V_T^*  & 0 & 0
                \\[.2cm]
                0 & 0 & \Gamma_{n+}  & 0
                \\[.2cm]
                0 & 0 & 0 & 0
	\ea
	\right],
\end{eqnarray}
and
\begin{eqnarray}
\non
		{\bf M}_{n,0}
		~=~
		\left[
                \ba{cccc}
                \Omega_n & 0 & 0 & I
                \\[.2cm]
                0 & \Omega_n & 0 & 0
                \\[.2cm]
                0 & 0 & \Omega_n & 0
                \\[.2cm]
                I & 0 & 0 & 0
                \ea
                \right],
                \quad
                {\bf M}_{n,-1}
                ~=~
		\left[
		\ba{cccc}
                U_T & 0 & 0 & 0
                \\[.2cm]
                0 & U_T & 0 & 0
                \\[.2cm]
                0 & 0 & U_T & 0
                \\[.2cm]
                0 & 0 & 0 & 0
                \ea
		\right],
		\quad
		{\bf M}_{n,+1}
		~=~
		\left[
		\ba{cccc}
                U_T^* & 0 & 0 & 0
                \\[.2cm]
                0 & U_T^* & 0 & 0
                \\[.2cm]
                0 & 0 & U_T^* & 0
                \\[.2cm]
                0 & 0 & 0 & 0
                \ea
		\right],
\end{eqnarray}
where
\begin{eqnarray*}
	\Gamma_n
	&=&
	-\dfrac{1}{Re} \left( \partial_{yy} \,-\, (((n\,+\,0.5)\alpha_r \,+\, \gamma)^2 \,+\, \beta^2) \right)
	\,+\,
	\mri \left((n\,+\, 0.5)\alpha_r+\gamma\right)U_B ~-
	\\[.15cm]
	&&
	~\mri (n\,+\, 0.5)\alpha_rc \,+\, V_B\partial_{y},
	\\[0.25cm]
	\Gamma_{n-}
	&=&
	\mri \left((n \,+\, 1.5)\alpha_r \,+\, \gamma \right) U_T \;+\; V_T\,\partial_{y},
	\\[0.35cm]
	\Gamma_{n+}
	&\!\!\!=\!\!\!&
	\mri \left((n \,-\, 0.5)\alpha_r \,+\, \gamma \right) U_T^* \;+\;V_T^*\,\partial_{y},
	\\[0.25cm]
	\Omega_n
	&=&
	U_B \;-\; \dfrac{2\,\mri }{Re} \left((n \,+\, 0.5)\alpha_r \,+\, \gamma \right).
\end{eqnarray*}
\end{subequations}

\section{Operators for streamwise streaks}
\label{sec.appendix1}

For the case study considered in Sec.~\ref{sec.Nlstreak}, the operators ${\bf{L}}_{n,m}$ and ${\bf{M}}_{n,m}$ in ${\bf{L}}_F$ and ${\bf{M}}_F$ from Eq.~\eqref{eq.PFE} are of the form:
\begin{subequations}
	\begin{eqnarray}
		{\bf L}_{n,0}
		&=&
		\left[
		\ba{cccc}
                        \Gamma_n \,-\, \partial_y\, V_B & \partial_y\, U_B & 0 & \mri \, \alpha
                        \\[.2cm]
                        0 & \Gamma_n \,+\, \partial_y\, V_B & 0 & \partial_y
                        \\[.2cm]
                        0 & 0 & \Gamma_n & \mri\, n\,\beta
                        \\[.2cm]
                        \mri\, \alpha & \partial_y & \mri\, n\,\beta & 0
		\ea
	\right],
	\label{eq.appendix-Ln0}
	\\[.35cm]
	\non
	{\bf L}_{n,-1}
	&=&
	\left[
	\ba{cccc}
                \mri\, \alpha\,  {U_{S,1}} & \partial_y\,  {U_{S,1}} & \mri\, \beta\,  {U_{S,1}} & 0
                \\[.2cm]
                0 & \mri\, \alpha\, {U_{S,1}}  & 0 & 0
                \\[.2cm]
                0 & 0 & \mri\, \alpha\,  {U_{S,1}}  & 0
                \\[.2cm]
                0 & 0 & 0 & 0
	\ea
	\right],
	\quad \quad
	{\bf L}_{n,+1}
	~=~
	\left[
	\ba{cccc}
                \mri\, \alpha\,  {U_{S,1}^*} & \partial_y\,  {U_{S,1}^*} & -\mri\, \beta\,  {U_{S,1}^*} & 0
                \\[.2cm]
                0 & \mri\, \alpha\,  {U_{S,1}^*}  & 0 & 0
                \\[.2cm]
                0 & 0 & \mri\, \alpha\,  {U_{S,1}^*}  & 0
                \\[.2cm]
                0 & 0 & 0 & 0
	\ea
	\right],
	\end{eqnarray}
and
	\begin{eqnarray}
	 {{\bf M}_{n,0}}
	& {=} &
	 {
	\left[
        \ba{cccc}
                U_B - \dfrac{2\,\mri\, \alpha}{Re} & 0 & 0 & {0}
                \\[.2cm]
                0 & U_B - \dfrac{2\,\mri\, \alpha}{Re} & 0 & 0
                \\[.2cm]
                0 & 0 & U_B - \dfrac{2\,\mri\, \alpha}{Re} & 0
                \\[.2cm]
                I & 0 & 0 & 0
        \ea
        \right],
        }
        \label{eq.appendix-Mn0}
	\\[.35cm]
	\non
	{\bf M}_{n,-1}
	&=&
	\left[
	\ba{cccc}
                 {U_{S,1}} & 0 & 0 & 0
                \\[.2cm]
                0 &  {U_{S,1}} & 0 & 0
                \\[.2cm]
                0 & 0 &  {U_{S,1}} & 0
                \\[.2cm]
                0 & 0 & 0 & 0
	\ea
	\right],
        \quad \quad
        {\bf M}_{n,+1}
        ~=~
	\left[
	\ba{cccc}
                 {U_{S,1}^*} & 0 & 0 & 0
                \\[.2cm]
                0 &  {U_{S,1}^*} & 0 & 0
                \\[.2cm]
                0 & 0 &  {U_{S,1}^*} & 0
                \\[.2cm]
                0 & 0 & 0 & 0
	\ea
	\right],
	\end{eqnarray}
where
\begin{align*}
	\Gamma_n
	~=~
	-\dfrac{1}{Re}
	\left( \partial_{yy} \,-\, \alpha^2
	\,+\,
	(n\, \beta)^2
	\right)
	\;+\;
	\left(
	-\mri\, \omega \,+\, \mri \, \alpha\, U_B \,+\, V_B\,\partial_{y}
	\right).
\end{align*}	
\end{subequations}
Note that consistent with nonlinear PSE, the boundary conditions on the MFD require special treatment; see~\cite{berherspa92}. Following~\cite{haj94,limal96,daymanrey01} which showed that the streamwise pressure gradient is the main contributor to the residual ellipticity in the PSE, we remove the pressure gradient from the streamwise velocity momentum (${\bf M}_{n,0}(1, 4)=0$) to ensure a well-posed streamwise march.

In subsequent PFE runs, the second harmonic and MFD also augment the base flow (cf.~Eq.~\eqref{eq.streakbase-rerun}) and $U_B$ and $V_B$ in Eqs.~\eqref{eq.appendix-Ln0} and~\eqref{eq.appendix-Mn0} denote the combination of the Blasius profile and the MFD. Interactions with the second harmonic generated from previous PFE runs are facilitated by the off-diagonal operators:
\begin{subequations}
	\begin{eqnarray*}
	{\bf L}_{n,-2}
	&=&
	\left[
	\ba{cccc}
                \mri\, \alpha\, U_{S,2} & \partial_y\, U_{S,2} & 2\mri\, \beta\, U_{S,2} & 0
                \\[.2cm]
                0 & \mri\, \alpha\, U_{S,2}  & 0 & 0
                \\[.2cm]
                0 & 0 & \mri\, \alpha\, U_{S,2}  & 0
                \\[.2cm]
                0 & 0 & 0 & 0
	\ea
	\right],
	\quad \quad
	{\bf L}_{n,+2}
	~=~
	\left[
	\ba{cccc}
                \mri\, \alpha\, U_{S,2}^* & \partial_y\, U_{S,2}^* & -2\mri\, \beta\, U_{S,2}^* & 0
                \\[.2cm]
                0 & \mri\, \alpha\, U_{S,2}^*  & 0 & 0
                \\[.2cm]
                0 & 0 & \mri\, \alpha\, U_{S,2}^*  & 0
                \\[.2cm]
                0 & 0 & 0 & 0
	\ea
	\right],
	\end{eqnarray*}
and
\begin{eqnarray*}
	{\bf M}_{n,-2}
	&=&
	\left[
	\ba{cccc}
                U_{S,2} & 0 & 0 & 0
                \\[.2cm]
                0 & U_{S,2} & 0 & 0
                \\[.2cm]
                0 & 0 & U_{S,2} & 0
                \\[.2cm]
                0 & 0 & 0 & 0
	\ea
	\right],
        \quad \quad
        {\bf M}_{n,+2}
        ~=~
	\left[
	\ba{cccc}
                U_{S,2}^* & 0 & 0 & 0
                \\[.2cm]
                0 & U_{S,2}^* & 0 & 0
                \\[.2cm]
                0 & 0 & U_{S,2}^* & 0
                \\[.2cm]
                0 & 0 & 0 & 0
	\ea
	\right].
	\end{eqnarray*}
\end{subequations}

\section{Grid-convergence and dependence on the number of harmonics}
\label{sec.convergence}

We examine the influence of the wall-normal grid-resolution ($N_y$) and the number of harmonics ($N$) considered in the PFE progression on the convergence of results obtained in Secs.~\ref{sec.H-type-transition} and~\ref{sec.Nlstreak}. To quantify convergence, the kinetic energy of the most important mode, i.e., the $(1, 1)$-subharmonic mode in H-type transition and the MFD of streaks are computed at various streamwise locations and stored in the vector $E$. The total energy in Tables~\ref{table-conv-htype} and~\ref{table-conv-streaks} denotes the aggregate kinetic energy in the streamwise direction and is computed using the Euclidean norm of the vector $E$, i.e., $\|E\|_2$. To quantify grid-independence, we use the relative error $\|E-E_r\|_2/\|E_r\|_2$, where $E_r$ is the kinetic energy obtained by refining resolution (in $N_y$ or in the number of harmonics).

\begin{table}[htb!]
\tabcolsep 0pt \caption{Convergence of results in the study of H-type transition}
\begin{center}
{\rule{0.95\textwidth}{1pt}}
\begin{tabular*}{.95\textwidth}{@{\extracolsep{\fill}} cccc}
  number of harmonics ($N$) & number of collocation points ($N_y$) & total energy of $(1,1)$ mode & relative error (\%) \\
  \hline \\[-0.2cm]
  2 & 40 & $0.15818$ & $15.6$\\[0.05cm]
  2 & 80 & $0.18266$ & $0.54$\\[0.05cm]
  2 & 160 & $0.18366$ & $\cdots$ \\[0.05cm]
  \hline \\[-0.2cm]
  2 & 80 & $0.18266$ & $0.31$\\[0.05cm]
  3 & 80 & $0.18209$ & $0.0008$\\[0.05cm]
  4 & 80 & $0.18209$ & $\cdots$\\[0.05cm]
  \hline
\end{tabular*}
\end{center}
\label{table-conv-htype}
\end{table}

\begin{table}[htb!]
\tabcolsep 0pt \caption{Convergence of results in the study of laminar streaks}
\begin{center}
{\rule{0.95\textwidth}{1pt}}
\begin{tabular*}{.95\textwidth}{@{\extracolsep{\fill}} cccc}
  number of harmonics ($N$) & number of collocation points ($N_y$) & total energy of MFD mode & relative error (\%) \\
  \hline \\[-0.2cm]
  3 & 40 & $1.5375\times10^{-3}$ & $0.6$\\[0.05cm]
  3 & 80 & $1.5471\times10^{-3}$ & $0.002$\\[0.05cm]
  3 & 160 & $1.5471\times10^{-3}$ & $\cdots$\\[0.05cm]
  \hline \\[-0.2cm]
  3 & 80 & $1.5471\times10^{-3}$ & $0.01$\\[0.05cm]
  4 & 80 & $1.5473\times10^{-3}$ & $0.005$\\[0.05cm]
  5 & 80 & $1.5472\times10^{-3}$ & $\cdots$\\[0.05cm]
  \hline
\end{tabular*}
\end{center}
\label{table-conv-streaks}
\end{table}

	\newpage

\end{document}